\documentclass{article}

\usepackage{arxiv}

\usepackage[utf8]{inputenc} 
\usepackage[T1]{fontenc}    
\usepackage{hyperref}       
\usepackage{url}            
\usepackage{graphicx}
\usepackage{epstopdf}
\usepackage{amsfonts}
\usepackage{amsmath,amssymb} 
\usepackage{amsbsy}
\usepackage{cite}
\usepackage{algorithm}
\usepackage{algpseudocode}
\usepackage{caption}
\usepackage{caption}
\usepackage{subcaption}
\usepackage{titlesec}
\usepackage{gensymb}
\usepackage{xcolor}
\usepackage{hyperref}
\usepackage{paralist}
\hypersetup{
    colorlinks=true,
    linkcolor=blue,
    filecolor=magenta,      
    urlcolor=cyan,
}

\title{Causal coupling inference from multivariate time series based on ordinal partition transition networks}

\author{
 Narayan Puthanmadam Subramaniyam \\
 Faculty of Medicine and Health Technology \\
 Tampere University \\
 \texttt{narayan.subramaniyam@tuni.fi} \\
   \And
           Reik V. Donner \\
              Magdeburg--Stendal University of Applied Sciences \emph{and} \\
              Potsdam Institute for Climate Impact Research (PIK) -- Member of the Leibniz Association\\
  \And
            Davide Caron \\
            Enhanced Regenerative Medicine \\
            Istituto Italiano di Tecnologia\\
\And
            Gabriella Pannucio \\
            Enhanced Regenerative Medicine \\
            Istituto Italiano di Tecnologia \\
            
\And    
 Jari Hyttinen \\
            Faculty of Medicine and Health Technology \\
            Tampere University \\

}

\begin{document}
\maketitle
\begin{abstract}
Identifying causal relationships is a challenging yet crucial problem in many fields of science like epidemiology, climatology, ecology, genomics, economics and neuroscience, to mention only a few. Recent studies have demonstrated that ordinal partition transition networks (OPTNs) allow inferring the coupling direction between two dynamical systems. In this work, we generalize this concept to the study of the interactions among multiple dynamical systems and we propose a new method to detect causality in multivariate observational data. By applying this method to numerical simulations of coupled linear stochastic processes as well as two examples of interacting nonlinear dynamical systems (coupled Lorenz systems and a network of neural mass models), we demonstrate that our approach can reliably identify the direction of interactions and the associated coupling delays. Finally, we study real-world observational microelectrode array electrophysiology data from rodent brain slices to identify the causal coupling structures underlying epileptiform activity. Our results, both from simulations and real-world data, suggest that OPTNs can provide a complementary and robust approach to infer causal effect networks from multivariate observational data. 


\end{abstract}

\section{Introduction}
\label{intro}
The detection of causal interactions is a fundamental problem in both natural and social sciences \cite{Ref1, Ref26}. Reliable statistical inference of causality can aid in making predictions, which may eventually allow to design proper intervention strategies \cite{Ref2}. For example, in neurological disorders such as epilepsy, predicting a seizure from electroencephalography (EEG) recordings and preventing its occurrence is a long-standing problem that has not yet been completely solved. In this context, reliable seizure prediction algorithms and consequent appropriate interventions to prevent seizure occurrence could be life-saving for the patient. 

In the past few decades, several approaches have been developed to both identify and quantify the interdependence between observational time series. The Granger causality test can be used to infer causality between two time series \cite{Ref3}, and its extension, partial directed coherence, allows to infer causality from multivariate data \cite{Ref4}. The Granger causality test is based on linear regression. Under this framework, a causal relation of one system (or variable) $X$ affecting another system $Y$ (i.e., $X \rightarrow Y$) is statistically inferred if the variance of the prediction error of the behavior of $Y$ can be reduced by including past information about observations of $X$ in the regression model for $Y$. Since the classical Granger causality analysis is model-based and (in a strict sense) only valid for linear systems, more general bivariate approaches based on information theory have been proposed for identifying causality in applications to nonlinear dynamical systems \cite{Ref7}. These methods include, among others, transfer entropy \cite{Ref5}, time-delayed mutual information \cite{Ref6}, and the multivariate extension of transfer entropy \cite{Ref7}. In particular, transfer entropy can be considered as a generalization of Granger causality for nonlinear systems \cite{Ref8}, while it has been shown to be equivalent to Granger causality for linear Gaussian models \cite{Ref9}.  

With the rising availability of powerful computer infrastructures and large observational data sets, the study of statistical interdependencies based on multivariate time series has found widespread applications, for instance, in the area of neuroscience. In the context of multi-channel EEG recordings, such interdependencies can be related to the concept of functional connectivity, which commonly refers to the statistical associations between pairs of signals that are measured in terms of simple linear correlations or variants thereof and, thus, cannot distinguish between direct and indirect connectivity \cite{Ref32}. Several statistical association measures have been employed to estimate functional connectivity from EEG data, including Pearson correlation, mutual information, synchronization likelihood, and phase locking value, all of which are symmetric measures and do not provide information regarding the direction of the associated information flow \cite{Ref35,Ref36,Ref37,Ref38,Ref6}. 

While studying complex systems such as neuronal networks, it is often important to identify not only the symmetric statistical associations, but also the causal relationships (i.e. driver-response relationships) between the involved sub-systems \cite{Ref26}. For example, effective connectivity \cite{Feldt2011} between individual neurons (or ensembles of neurons) is characterized by direct (causal) relationships. Hence, statistical inference of causality based on time series is a more informative approach to better understand the interplay between neuronal connectivity and dynamics, and is the key to study the structure-function relationship of neuronal networks \cite{Ref33}. To estimate effective connectivity, recent works using Granger causality based methods such as directed coherence and transfer entropy have been proposed \cite{Ref39}. Although these concepts provide asymmetric measures, they cannot distinguish between causal (direct) and non-causal (indirect) interdependence among two subsystems. By contrast, partial directed coherence \cite{Ref4} (which has been often used in EEG studies) can distinguish between direct and indirect causal links but assumes a simple linear multivariate autoregressive model for describing the data. More recently, Bayesian filtering approaches have been suggested to estimate connectivity. However, such approaches also make strong assumptions about the underlying dynamical model \cite{Ref45, Ref46, Ref47}. Another approach known as dynamic causal modeling \cite{Ref40} has also been widely used to estimate causality among brain regions using EEG, magnetoencephalography or functional magnetic resonance imaging measurements. However, dynamic causal modeling is also a model-based approach and makes strong assumptions on the process by which the data are generated. In addition, it requires the pre-specification of several competing hypotheses, which may not always be available. For a detailed review on various methods for estimating neural connectivity and their applications in neuroscience, we further refer to \cite{Ref41, Ref42} and the references therein.

In the last years, a great deal of interest has emerged in characterizing dynamical systems using complex network based time series analysis methods \cite{Ref10}. Those methods include, among others, recurrence networks \cite{Ref11, Ref27, Ref28}, visibility graphs, \cite{Ref12} and transition networks \cite{Ref13}, all of which feature different definitions of nodes and links in the resulting network representations of the time series under study. For instance, in the case of recurrence networks, the edges are defined based on the proximity of observed states in phase space, whereas for visibility graphs \cite{Ref11}, mutually visible elements in a univariate time series are linked to form a network \cite{Ref12}. Finally, in the case of transition networks \cite{Ref13}, certain discrete states or patterns are defined as nodes, and if one of these dynamical structures is followed by the other with nonzero probability along the observed (or reconstructed) trajectory, a directed edge is established between the corresponding nodes \cite{Ref14}. However, the majority of previous applications of the aforementioned methods have focused on univariate time series, while a generalization to multivariate time series would be a necessary step to allow detecting signatures of causality. This aspect has not yet been systematically explored in full detail in the recent literature. 

In this work, we consider a particular class of transition networks known as ordinal partition transition networks (OPTNs) \cite{Ref30} to infer causality from multivariate time series. OPTNs are based on the ordinal patterns embedded in a time series, the systematic analysis of which has originally been proposed in \cite{Ref29}. Such ordinal patterns reflect the respective rank order among a predefined sequence of univariate observation values. Identifying the series of subsequent ordinal patterns in a univariate time series results in a particular symbolic representation of the observed system's trajectory. It has been shown that under specific conditions, this ordinal partition exhibits the generating property, which makes it attractive for applications since it implies topological conjugacy between phase space and the ordinal symbolic dynamics \cite{Ref13}. For an unconstrained stochastic process, all possible ordinal patterns occur with equal probability. By contrast, for a time series produced by deterministic dynamics, certain ordinal patterns commonly do not appear, which are known as forbidden patterns \cite{Ref17} and provide a window of opportunity to test for possible determinism in a time series \cite{Amigo2008,Amigo2010,Kulp2016a}. Furthermore, given the symbolic representation of the phase space trajectory (which controls the respective frequency of the different ordinal patterns), it is possible to compute a variety of dynamical characteristics, such as permutation entropy or a plethora of statistical complexity measures that can be defined based on the latter concept. Taken together, ordinal pattern--based analysis offers several advantages as compared to other more traditional nonlinear time series analysis techniques. The resulting methods are conceptually simple, computationally fast, and can capture information about the short-range temporal structure of the underlying time series \cite{Ref43}. In addition, they have been shown to be robust against additive noise. Finally, the calculation of ordinal patterns does not require any \textit{a priori} knowledge of the data range, which is practical and advantageous in time series analysis \cite{Ref43}.

An OPTN is based on the ordinal symbolic encoding of a time series and consists of \cite{Ref14} 
\begin{inparaenum}[(i)]
\item nodes, which represent the individual ordinal patterns and
\item probability-weighted edges, which represent the transition frequencies between two successive ordinal patterns.
\end{inparaenum}
Previous applications of statistical complexity measures derived from OPTNs include the classification of cardiac dynamics based on electrocardiography data \cite{Ref13, Ref15} and the analysis of EEG data from healthy and epileptic humans \cite{Ref16}. Although recent attempts have addressed bi- and multivariate extensions of OPTNs \cite{Ref17}, often with the aim of characterizing different types of synchronization transitions, they have not yet provided thorough information about causal relationships among multivariate time series. Most recently, Ruan \emph{et~al.} \cite{Ref14} have proposed a strategy for the estimation of several complexity measures based upon bipartite OPTNs, which allows for statistical inference of the coupling direction among paired time series. However, their approach is limited by its bivariate nature. Hence, when applied to a multivariate data set, it cannot distinguish between direct and indirect causal connections among the individual time series. 

To overcome the aforementioned limitations of previous OPTN--based methods, in this work we propose an extension of OPTN--based time series analysis, which leverages the construction of multiple bipartite OPTNs (M-OPTN) to account for multivariate (i.e., comprising more than two components) time series. Specifically, we outline and thoroughly test an approach for distinguishing direct from indirect causal connections based on the conditional Shannon entropies of the bipartite constituents of the M-OPTN \cite{Ref14}. In order to demonstrate the effectiveness of our approach, we apply the proposed method to coupled linear stochastic processes, nonlinear dynamical systems (exemplified by three interacting Lorenz systems), and a network of coupled neural mass models. The latter type of system has been shown previously to mimic the dynamics exhibited by neurophysiological time series \cite{Ref22} and thus serves as a validation tool to test the applicability of our method to neuronal time series. Finally, as a real-world example, we study \textit{in vitro} microelectrode array (MEA) recordings from an \textit{in vitro} model of acute ictogenesis, i.e., rodent brain slices in which epileptiform discharges are induced by pharmacological treatment. The network interactions and the associated delays of epileptiform discharges propagation that occur in this \textit{in vitro} model have been extensively characterized \cite{Ref21, Ref24} and serve as a reliable reference to validate the causal network relationships and delays estimated by the proposed method. 

\section{Methodology}
\label{sec:1}
\subsection{Ordinal partition transition networks}
Given a univariate time series $X = \{x_t\}_{t=1}^{T}$, following Takens' embedding theorem, we can qualitatively reconstruct the underlying phase space trajectory by using $M$ successively lagged replications of $X$, each separated by a lag $d$, yielding the vector
\begin{equation}
    \mathbf{z}_t = [x_{t}, x_{t+d}, \ldots , x_{t+(M-1)d}],
    \label{takens_eqn}
\end{equation}
 for $t=1$ to $T-(M-1)d$, where $M$ and $d$ are the embedding dimension and delay, respectively. Each embedding vector $\mathbf{z}_t$ is mapped to a sequence of integers $(s_0, s_1, \ldots, s_{M-1})$ that describes the rank order of its components (with $0$ indicating the smallest value) and is a unique permutation of the set $\{0,1,\ldots, M-1\}$, thereby satisfying

\begin{equation}
x_{t+s_{0}d} \le x_{t+s_{1}d} \le x_{t+s_{2}d} \le \ldots \le x_{t+s_{M-1}d}   
\label{op_2}
\end{equation}
and
\begin{equation}
    s_{l-1} < s_l \quad \text{if} \quad x_{t-s_{l-1}} = x_{t-s_{l}}.
    \label{op_3}
\end{equation}
Note that there exist $M!$ different possible ordinal patterns when a time series is embedded in $M$ dimensions, and we denote these patterns by $\pi_{1}, \pi_{2}, \ldots, \pi_{M!}$. 

 As an example, consider a $5$-dimensional embedding of a time series yielding an embedding vector
 \begin{equation}
 \{x_t, x_{t+d}, x_{t+2d}, x_{t+3d}, x_{t+4d}\} = \{ 3, 9, 10, 1, 6\}.
 \end{equation}
 Here, $x_{t+3d} < x_{t} < x_{t+4d} < x_{t+d} < x_{t+2d}$, and thus this partition would be mapped to the ordinal pattern or symbol $\pi_{k} = \{3, 0, 4, 1, 2\}$. The exact numerical value of the resulting integer index $k \in \{1, \ldots, M!\}$ depends on the specific sorting of the permutations, the default of which may differ among different algorithms and programming languages.
 
 For a univariate time series, we can then construct an (unweighted or weighted) OPTN with $M!$ nodes by first repeating this encoding procedure for each embedding vector. A weighted OPTN is obtained by setting the weight of the edge between two nodes (permutations) to be equal to the empirical frequency of ``transitions'' (i.e., successive occurrences) between the corresponding possible ordinal patterns. An unweighted OPTN simply contains a directed edge of unit weight between the corresponding nodes if this frequency is nonzero. 

Ruan \emph{et~al.} recently extended the idea of OPTNs to bivariate time series, that may be interacting either linearly or non-linearly \cite{Ref14}. In their framework, given the time series $\{x_{1,t}\}_{t=1}^{T}$ and $\{x_{2,t}\}_{t=1}^{T}$, derived from two dynamical systems $X_1$ and $X_2$, we can derive the associated sequences of ordinal patterns underlying each time series as described above, containing the ordinal patterns $\pi_i^{x_1}$ and $\pi_j^{x_2}$ for $X_1$ and $X_2$, respectively. One can now compute the (instantaneous or time-lagged) conditional co-occurrence frequencies $p(\pi_j^{x_2} | \pi_i^{{x_1},\tau})$ by simply counting the number of cases in which $\pi_j^{x_2}$ occurs with a time-lag of $\tau$ following an occurrence of $\pi_i^{x_1}$. Note that we deviate here from the notation in \cite{Ref14} in order to allow a straightforward generalization of the concept of co-occurrence probabilities by including multiple variables $X_i$ that may possibly cause variations of the given $X_2$ at different delays $\tau_i$ (see Section 2.2 below).Thus, $\tau = 0$ corresponds to looking at simultaneous co-occurrence while $\tau > 0$ would imply looking at lagged co-occurrence. Given these conditional co-occurrence frequencies, Ruan \emph{et~al.} proposed the estimation of the conditional Shannon entropy \cite{Ref14} (in the following referred to as co-occurrence entropy (CE)), given as

\begin{equation}
    H_{\tau} (X_2 | X_1) = -\sum_{i=1}^{M!} \sum_{j=1}^{M!} p(\pi_i^{x_1, \tau}, \pi_j^{x_2}) \log_2 p(\pi_j^{x_2} | \pi_i^{x_1,\tau}),
    \label{CE}
\end{equation}
\noindent
which gives the interaction $X_1 \rightarrow X_2$ at a delay of $\tau$. The interaction in the other direction at lag $\tau$, $X_2 \rightarrow X_1$ can be defined in an analogous way as follows,

\begin{equation}
    H_{\tau} (X_1 | X_2) = -\sum_{i=1}^{M!} \sum_{j=1}^{M!} p(\pi_j^{{x_2},\tau}, \pi_i^{x_1}) \log_2 p(\pi_i^{x_1} | \pi_j^{{x_2},\tau}),
\end{equation}

If $X_1$ and $X_2$ are independent and their different ordinal patterns uniformly distributed, then $p(\pi_j^{{x_2},\tau}, \pi_i^{x_1}) = \frac{1}{(M!)^2}$ and $p(\pi_i^{x_1} | \pi_j^{{x_2},\tau}) = \frac{1}{M!}$ and thus $H_{\tau} (X_1 | X_2) = \log_2 M!$, which is the upper bound for the conditional entropy value, denoted as $H_{max}$. On the other hand, if $X_1$ and $X_2$ are fully dependent, then $p(\pi_i^{x_1} | \pi_j^{{x_2},\tau}) = 1$ and ideally $H_{\tau} (X_1 | X_2) = 0$. Thus, as the strength of causal interaction from $X_2$ to $X_1$ at a given time lag $\tau$ increases, $H_{\tau} (X_1 | X_2)$ decreases.

\subsection{Causal inference strategy based on entropy measures from OPTNs}
\begin{figure}[h]
    \centering
        \includegraphics[scale=0.1]{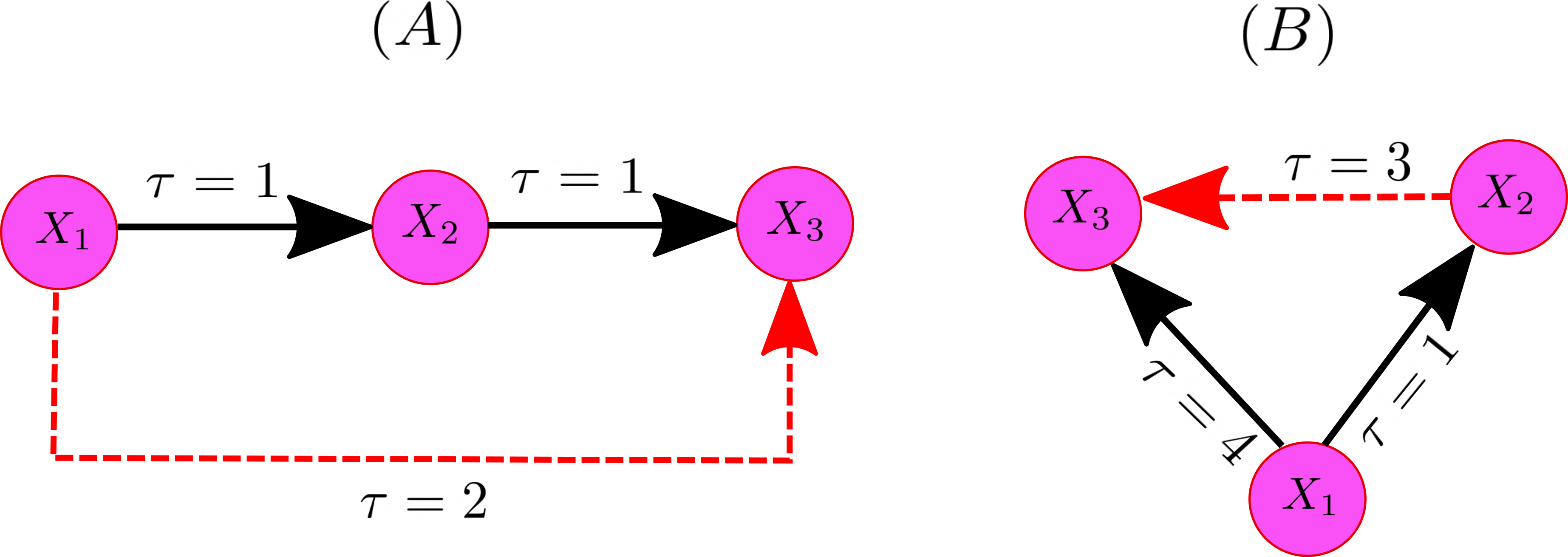}
    \caption{Example of spurious connections (shown as dotted red arrows) that would be inferred from a directed chain (A) and a fork pattern (B) if direct links are not distinguished from indirect links.}
    \label{example_networks}
\end{figure}

\begin{figure}[h]
    \centering
        \includegraphics[scale=0.3]{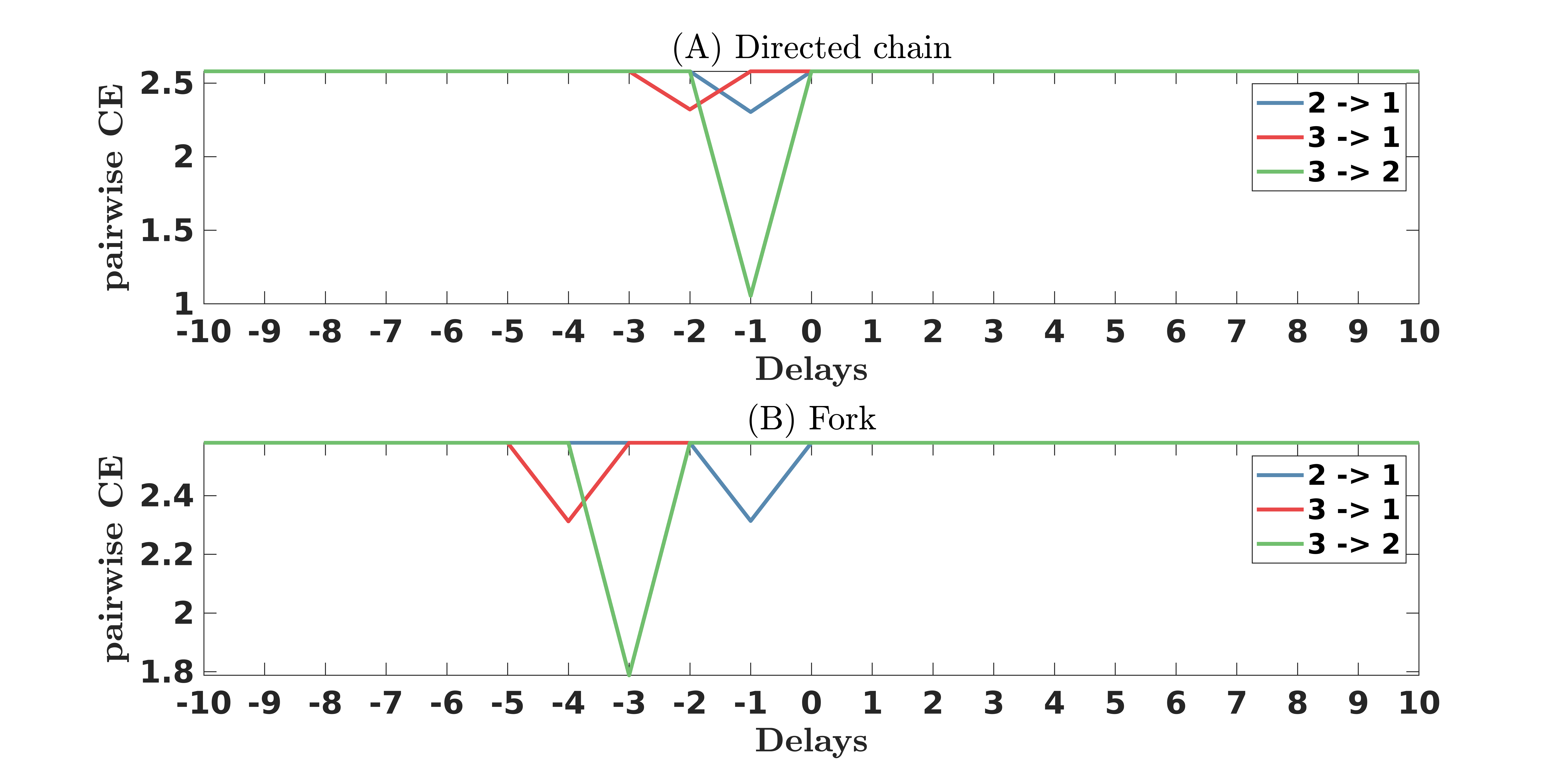}
    \caption{Causal network inferred inferred with the bivariate approach using OPTN--based CE for the directed chain (A) and fork (B) with a spurious connection $1\rightarrow2$ at $\tau=2$ for the directed chain and $2\rightarrow3$ at $\tau=3$ for the fork pattern.} 
    \label{ex_caual_biv}
\end{figure}

When dealing with multivariate data, i.e. data from three or more interacting systems, it is necessary to distinguish direct links from indirect ones. For example, consider a transitive chain $X_1 \rightarrow X_2 \rightarrow X_3$ (Figure \ref{example_networks} (A)). Applying the bivariate OPTN derived entropy measure as described in \cite{Ref14} would lead to the detection of a non-existing connection from $X_1 \rightarrow X_3$, as shown in Figure \ref{ex_caual_biv} (A). Similarly, consider a fork pattern, where $X_1 \rightarrow X_2$ and $X_1 \rightarrow X_3$ (Figure \ref{example_networks} (B)). Here, $X_1$ is a common driver to both $X_2$ and $X_3$ and this will lead to a spurious connection between the $X_2$ and $X_3$ as shown in Figure \ref{ex_caual_biv} (B). Note that in Figure \ref{ex_caual_biv}, an interaction $m \rightarrow n$ at negative delay is to be interpreted as $n \rightarrow m$. 

\begin{figure}[h]
    \centering
        \includegraphics[clip, trim=0.5cm 10cm 0.5cm 10cm, width=1.0\textwidth]{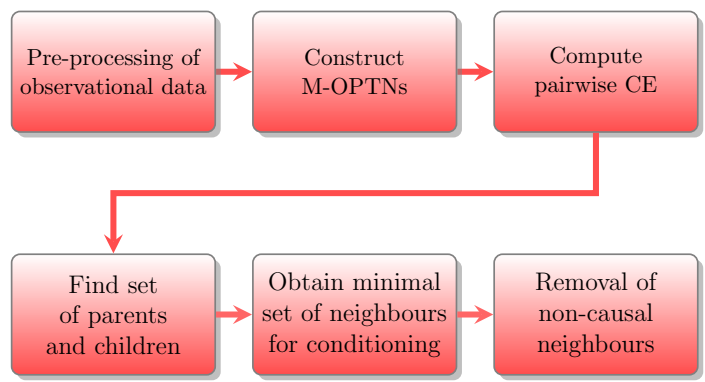}
    \caption{Proposed methodology}
    \label{fig_method}
\end{figure}

In order to distinguish such direct from indirect links, a sophisticated way of conditioning has to be employed, thereby generalizing the previous strictly bivariate approach. In the following, we will detail a possible methodology to use OPTN-- based entropy measures to infer causality from multivariate time series as outlined in Figure \ref{fig_method}. Note that this methodology rests upon certain general assumptions common to causal inference methods, most notably, the completeness of the set of variables analyzed (i.e. the absence of any possible hidden drivers).

\subsubsection*{STEP I$\colon$ Pre-processing of observational data}

Depending on the particular research problem, observational time series first have to be pre-processed, which includes standard procedures like band-pass filtering, resampling and removal of any noise or artifacts if present. If necessary, the data can be further divided into a number of overlapping windows to obtain a time-dependent estimate of the coupling measure (see below).

\subsubsection*{STEP II$\colon$ Construction of multiple OPTNs}
Given a time series, its phase space is reconstructed following Takens' embedding theorem. The components of the resulting embedding vectors are then rank ordered to obtain the symbolic representation of the time series. For an $N$-channel multivariate time series, $N$ such OPTNs are constructed, which will be referred to as multiple OPTNs (M-OPTNs) in the following.

\begin{algorithm}
\caption{Construct M-OPTN from multivariate time series}\label{alg:const_m-optn_1}
\begin{algorithmic}[1]
\Procedure{ComputeM-OPTN}{}
\State \textbf{Input:} Multivariate time series $X_1, X_2, \ldots, X_N$, embedding dimension $M$, and lag $d$.
\State \textbf{Output:} M-OPTN $\boldsymbol{\Pi}$
\For {$n = 1$ to $N$}
\For {$t=1$ to $T-(M-1)d + 1$}
\State Map $\mathbf{z}_{n,t}$ to symbol $(s_0, \ldots, s_{M-1})$ using Equations \eqref{op_2} to \eqref{op_3}.
\State Assign ordinal patterns $\boldsymbol{\Pi}[t,n] = {\pi}_k^{X_n}$ based on the respective permutation for each $t$ and $n$.
\EndFor
\EndFor
\EndProcedure
\end{algorithmic}
\label{alg1}
\end{algorithm}

\subsubsection*{STEP III$\colon$ Conditional entropy from M-OPTN}
After constructing the M-OPTN, we compute the bivariate information theoretic measure of CE as given in Equation~\eqref{CE} for each pair of variables and define the matrix
\begin{equation}
\mathbf{H}_\tau = 
    \begin{bmatrix}
 H_{\tau} (X_1) &H_{\tau} (X_1 \vert X_2) &\cdots &H_{\tau} (X_1 \vert X_N) \\
 H_{\tau} (X_2 \vert X_1) &H_{\tau} (X_2) &\cdots &H_{\tau} (X_2 \vert X_N)\\ 
 \vdots  &\vdots &\ddots &\vdots\\ 
 H_{\tau} (X_N \vert X_1) &H_{\tau} (X_N \vert X_2) &\cdots &H_{\tau} (X_N)
\end{bmatrix},
\end{equation}
\noindent
where each off-diagonal term $H_\tau(X_n \vert X_m)$ represents a possible causal link from the $m-th$ time series to the $n-th$ time series, i.e., $X_m \rightarrow X_n$, at delay $\tau$. Note that in what follows, we will not make use of the diagonal elements of $\mathbf{H}_\tau$ (representing the classical Shannon entropies -- i.e. in our specific case the permutation entropies -- of the individual processes), so that they could be safely ignored or just put to zero.

Let $\mathbf{H} = \{\mathbf{H}_{\tau_1}, \mathbf{H}_{\tau_2}, \ldots, \mathbf{H}_{\tau_J}\}$ denote the CE matrices obtained from the M-OPTN for a range of $J$ delays $ \boldsymbol{\mathrm{T}} = \{ \tau_1, \ldots, \tau_{J} \}$. Each of the matrices $\mathbf{H}_{\tau}$ obtained above is next thresholded using a hard threshold to obtain a new matrix $\mathbf{\hat{H}}_{\tau}$ with elements 
 \begin{equation}
    \hat{H}_\tau(X_n \vert X_m) = h_{nm\tau} =
    \begin{cases}
      H_{max}, & \text{if}\ {H}_\tau(X_n \vert X_m) \geq \lambda H_{max} \\
      H_\tau(X_n \vert X_m), & \text{otherwise}
    \end{cases}
    \label{thr}
  \end{equation}
\noindent
where $H_{max} = \log_2 M!$ and $\lambda$ is usually set between $0.99$ and $1$. Due to the finite sample size, two independent processes will not have a CE value exactly equal to $H_{max}$. To account for this numerical issue, we allow for some tolerance by setting the parameter $\lambda$. After this step, only dominant neighbors (both causal and non-causal) are retained for each node. The resulting matrix $\mathbf{\hat{H}}_\tau$ represents a weighted, directed network of $N$ nodes, where the strength of the link between the node at delay $\tau$ is inversely proportional to the CE value $H_\tau(X_n \vert X_m)$ and no link exists between two nodes if $H_\tau(X_n \vert X_m) \geq \lambda H_{max}$.

\begin{algorithm}
\caption{Compute pairwise CE using M-OPTN}\label{alg:const_m-optn_2}
\begin{algorithmic}[1]
\Procedure{ComputeCE}{}
\State \textbf{Input:} M-OPTN $\boldsymbol{\Pi}$, threshold $\lambda$ and range of delays $ \boldsymbol{\mathrm{T}} = \{ \tau_1, \ldots, \tau_{J} \}$.
\State \textbf{Output:} CE matrix $\mathbf{\hat{H}}$
\For {$i=1$ to $J$}
\For {$n = 1$ to $N$}
\For {$m = 1$ to $N$}
\State  Compute $H_{\tau_i} (X_n | X_m)$ using Equation \eqref{CE}
\EndFor
\EndFor
\EndFor
\State Set $\mathbf{H} = [\mathbf{H}_{\tau_1}, \mathbf{H}_{\tau_2}, \ldots, \mathbf{H}_{\tau_J}]$.
\State Threshold $\mathbf{H}$ using $\lambda$ and $H_{max}$ as shown in Equation \eqref{thr}.
\EndProcedure
\end{algorithmic}
\label{alg1}
\end{algorithm}

\subsubsection*{STEP IV$\colon$ Find set of parents and children}
The set of matrices $\mathbf{\hat{H}} = \{\mathbf{\hat{H}}_{\tau_1}, \mathbf{\hat{H}}_{\tau_2}, \ldots, \mathbf{\hat{H}}_{\tau_J}\}
$ represents a weighted multi-layer network $G = \{V, E\}$, where $V = \{X_1, \ldots, X_N\}$ denotes a set of nodes (i.e. the different component processes) common to the different layers (which represent the different delays $\tau_j$), while $E=\{E_1,\ldots,E_J\}$ is a set of edges that will commonly differ among the layers. 
For every node $X_m$ in the multi-layer network defined by $\mathbf{\hat{H}}$, we identify a set of $k_m$ parents $\mathcal{P}_{X_m} = \{p_1, \ldots, p_k\}$ at delays $\{\tau_{p_1}, \ldots, \tau_{p_{k_m}}\}$ and $l_m$ children $\mathcal{C}_{X_m} = \{c_1, \ldots, c_l\}$ at delays $\{\tau_{c_1}, \ldots, \tau_{c_{l_m}}\}$, respectively, which can span across all possible layers. The set of parents for a node $X_m$ is given by 
\begin{equation}
    \mathcal{P}_{X_m} = \{ {h}_{mn\tau} \vert {h}_{mj\tau} < H_{max},\ n\neq m,\ \tau\in\boldsymbol{\mathrm{T}} \}
    \label{parent_set}
\end{equation}
\noindent
where $h_{mj\tau}$ describes the element in row $m$ and column $j$ of the single-layer adjacency matrix $\mathbf{\hat{H}}_\tau$, cf.\ Equation~\eqref{thr}. 
In a completely analogous way, we define the set of children of $X_n$ node as 
\begin{equation}
\mathcal{C}_{X_n} = \{ {h}_{mn\tau} \vert {h}_{in\tau} < H_{max},\ m\neq n,\ \tau\in\boldsymbol{\mathrm{T}} \}
\label{children_set}
\end{equation}
where $h_{in\tau}$ represents the element in row $i$ and column $n$ of $\mathbf{\hat{H}}_\tau$. By defining those two sets for all nodes (variables) $X_m$ ($X_n$), we collapse the information contained in the multi-layer adjacency matrix $\mathbf{\hat{H}}$ to the essential strong bivariate (time-lagged) linkages among the set of considered variables.

\begin{algorithm}
\caption{Find set of parents and children from pairwise CE matrix}\label{alg:const_m-optn_3}
\begin{algorithmic}[1]
\Procedure{FindPC}{}
\State \textbf{Input:} Set of matrices $\mathbf{H}$, $H_{max}$.
\State \textbf{Output:} $\mathcal{P}_{X_n}$ at delays $\{\tau_{p_1}, \ldots, \tau_{p_k}\}$ and $\mathcal{C}_{X_n}$ at delays $\{\tau_{c_1}, \ldots, \tau_{c_l}\}$ for each node $X_n$.
\For {$n = 1$ to $N$}
\State  Compute $\mathcal{P}_{X_n}$ and $\mathcal{C}_{X_n}$ using Equation \eqref{parent_set} and \eqref{children_set}.
\EndFor
\EndProcedure
\end{algorithmic}
\label{alg1}
\end{algorithm}

\subsubsection*{STEP V$\colon$ Identifying the minimal set of neighbors for conditioning}

Given at set of parents for node $X_m$, $\mathcal{P}_{X_m} = \{p_1, \ldots, p_k\}$  at delays $\{\tau_{p_1}, \ldots, \tau_{p_k}\}$, to test for a causal connection from node $X_n$ to node $X_m$, we seek to define a minimal conditioning set $\mathcal{P}_{X_m}^{min} \subset \mathcal{P}_{X_m}$, given as

\begin{equation}
    \mathcal{P}_{X_m}^{min} = \mathcal{P}_{X_m} \cap \mathcal{\widehat{C}}_{X_n} = \{p_1^\prime, \ldots, p_r^\prime\}
    \label{eq_pmin}
\end{equation}
\noindent 
where $\mathcal{\widehat{C}}_{X_n}$ is the set of children of node $X_n$, which does not include $X_m$, i.e. $\mathcal{\widehat{C}}_{X_n}  = \mathcal{C}_{X_n} \setminus X_m$. The set $\mathcal{\widehat{C}}_{X_n}$ will be an empty set if $\mathcal{C}_{X_n} = \{X_m\}$, i.e. the only child of node $X_n$ is node $X_m$, and in this case we set $\mathcal{P}_{X_m}^{min} = \{X_m\}$. Also, it is possible that $\mathcal{P}_{X_m}^{min}$ will be an empty set due to no common elements between $\mathcal{P}_{X_m}$ and $\mathcal{\widehat{C}}_{X_n}$. In this case, we set $\mathcal{P}_{X_m}^{min}= \mathcal{P}_{X_m} \cap \mathcal{P}_{X_n}$. If $\mathcal{P}_{X_m}^{min}$ is still an empty set, then we set $\mathcal{P}_{X_m}^{min} = \{X_m\}$. Note that in each of these cases, along with the conditioning set $\mathcal{P}_{X_m}^{min}$, we also obtain the corresponding delays, $\{\tau_1^\prime, \ldots, \tau_r^\prime\}$, with $\vert \mathcal{P}_{X_m}^{min}\vert = r $. In order to facilitate reliable computation of CE (see STEP VI) to exclude non-causal neighbors, in all our applications we restrict ourselves here to $r = 3$ if $r>4$, by choosing the three most dominant neighbors, based on their respective CE values.

\begin{algorithm}
\caption{Find minimal set of neighbors for conditioning}\label{alg:const_m-optn_4}
\begin{algorithmic}[1]
\Procedure{FindMinConditioningSet}{}
\State \textbf{Input:} $\mathcal{P}_{X_n}$ at delays $\{\tau_{p_1}, \cdots, \tau_{p_k}\}$ and $\mathcal{C}_{X_n}$ at delays $\{\tau_{c_1}, \cdots, \tau_{c_l}\}$ for each node $X_n$ and $r$.
\State \textbf{Output:} $\mathcal{P}_{X_m}^{min}$
\For {$m = 1$ to $M$}
\For {$n = 1$ to $N$}
\State  Set $\mathcal{\widehat{C}}_{X_n}  = \mathcal{C}_{X_n} \setminus X_m$
\If {$\mathcal{\widehat{C}}_{X_n} = \emptyset$}
\State  $\mathcal{C}_{X_n} = \{X_m\}$
\State $\mathcal{P}_{X_m}^{min} = \{X_m\}$ 
\State \textbf{break}
\EndIf
\State Set $\mathcal{P}_{X_m}^{min}$ according to Equation \eqref{eq_pmin}.
\If {$\mathcal{P}_{X_m}^{min} = \emptyset$}
\State $\mathcal{P}_{X_m}^{min}= \mathcal{P}_{X_m} \cap \mathcal{P}_{X_n}$
\EndIf
\If {$\mathcal{P}_{X_m}^{min} = \emptyset$}
\State $\mathcal{P}_{X_m}^{min} = \{X_m\}$
\EndIf
\EndFor
\EndFor
\If {$\vert \mathcal{P}_{X_m}^{min}\vert > r$}
\State  Set $\vert \mathcal{P}_{X_m}^{min}\vert = r$
\EndIf
\EndProcedure
\end{algorithmic}
\label{alg1}
\end{algorithm}

\subsubsection*{STEP VI$\colon$ Removal of non-causal neighbors by proper conditioning}
To check if $X_n$ is a truly causal parent to $X_m$, we compute 

\begin{equation}
    \epsilon_{X_n} = H(X_m | \mathcal{P}_{X_m}^{min}) - H(X_m | \mathcal{P}_{X_m}^{min}, X_n).
    \label{eq_epsilon}
\end{equation}
\noindent
If $X_n$ is an indirect causal connection to $X_m$, then $\epsilon_{X_n} = 0$, since conditioning on $X_n$ should not reduce $H(X_m | \mathcal{P}_{X_m}^{min})$ any further. However, since we are dealing with finite data, $\epsilon_{X_n} \approx 0$. In practice, we set another pre-defined threshold $\delta$ and if $\epsilon_{X_n} < \delta$, then $X_n$ is considered as an indirect causal link to $X_m$. The CE $H(X_m|\mathcal{P}_{X_m}^{min})$, with $\vert{\mathcal{P}_{X_m}^{min}\vert} = r$, is given as

\begin{equation}
    H(X_m|\mathcal{P}_{X_m}^{min})  = -\sum_{i=1}^{M!} \sum_{j=1}^{M!} p(\pi_i^{p_1^\prime,\tau_{1}^\prime}, \dots, \pi_i^{p_r^\prime, \tau_{r}^\prime}, \pi_j^{x_m})  \log p(\pi_j^{x_m} | \pi_i^{p_1^\prime,\tau_{1}^\prime}, \dots, \pi_i^{p_r^\prime, \tau_{r}^\prime}).
  \label{eq_cond_N}
\end{equation}
The CE $H(X_m|\mathcal{P}_{X_m}^{min}, X_n)$ is defined in an analogous way. 

\begin{algorithm}
\caption{Remove non-causal neighbors based on CE differences}\label{alg:const_m-optn_5}
\begin{algorithmic}[1]
\Procedure{RemoveNCN}{}
\State \textbf{Input:} Matrix $\mathbf{H}$, $\delta$.
\State \textbf{Output:} Matrix $\mathbf{H}$.
\For {$m = 1$ to $M$}
\For {$n = 1$ to $N$}
\State  Compute $\epsilon_{X_n}$ using Equations \eqref{eq_epsilon} and \eqref{eq_cond_N}.
\If {$\epsilon_{X_n} < \delta$}
\State  Set $H(X_m \vert X_n) = H_{max}$
\EndIf
\EndFor
\EndFor
\EndProcedure
\end{algorithmic}
\label{alg1}
\end{algorithm}

\section{Numerical examples}

In the following, we present results from the application of the proposed methodology to simulations of linearly interacting stochastic processes as well as interacting nonlinear dynamical systems including  Lorenz systems and a network of neural mass models. In the case of interacting stochastic processes and interacting Lorenz system, we varied $\delta$ from $0$ to $0.5$. We also added observational noise to the simulated data,

\begin{equation}
    y(t) = x(t) + e(t)
\end{equation}
where $e(t) \sim \beta \mathcal{N}(0,1)$, where $\beta$ is the noise level (NL), which is set at $0.1$, $0.2$ and $0.4$ times the standard deviation of original noise free time series. 

In the case of the network of neural mass models, we varied $\delta$ from $0$ to $0.25$ and added noise at the level of $0.5$ and $1$ times the standard deviation of original noise-free time series, which corresponds to realistic signal-to-noise ratios ($SNR$) commonly found in EEG data (Amplitude $SNR=2$ and $1$, respectively).

For evaluating the results of our methodology for the different simulations, we define the number of true positives (\#TP) as the number of correctly identified links, and the number of false negatives (\#FN) as the number of missed links. The number of false positives (\#FP) is defined as the number of incorrectly identified links, and the number of true negatives (\#TN) corresponds to the number of correctly identified non-links. We then define the true positive rate (TPR) and false positive rate (FPR) as
\begin{align}
\begin{split}
    TPR &= \frac{\#TP}{\#TP + \#FN}, \\
    FPR &= \frac{\#FP}{\#FP + \#TN}. 
\end{split}
\end{align}
\noindent
In addition, we use the $F_1$-score to quantify the accuracy of the method, which is given as
\begin{align}
    F_1 &= \frac{\#TP}{\#TP + 0.5(\#FP + \#FN)}.
\end{align}

\subsection{Interacting stochastic processes}
\label{sec:2}
\begin{figure}
    \centering
    \includegraphics[scale=0.5]{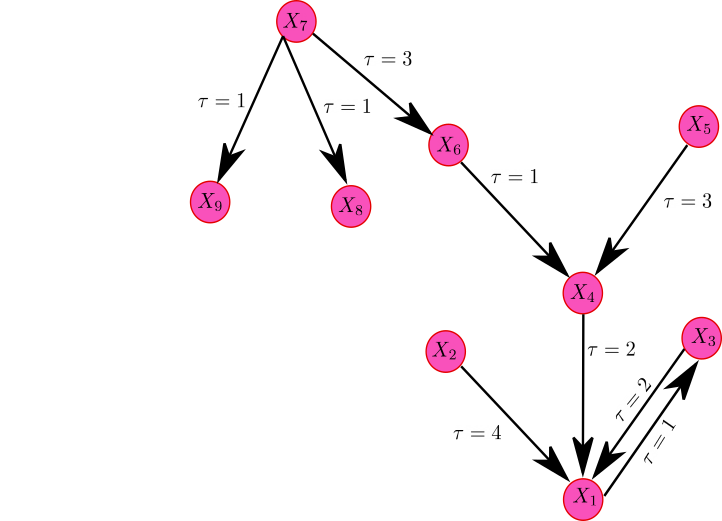}
    \caption{Multivariate autoregressive system described in Equation \eqref{MVAR}.}
    \label{stochastic}
\end{figure}

We first simulated the following multivariate autoregressive system such that it contains a directed chain as well as a fork, both of which can lead to spurious causal links:

\begin{eqnarray}
    x_{1,t} & = & 3.4x_{1,t-1}(1 - x_{1,t-1}^2e^{x_{1,t-1}^2}) + c_{21}x_{2,t-4} + c_{31}x_{3,t-2} + c_{41}x_{4,t-2} + 0.4u_{1,t}\nonumber \\
    x_{2,t} & = & 3.4x_{2,t-1}(1 - x_{2,t-1}^2e^{x_{2,t-1}^2}) + 0.4u_{2,t}\nonumber \\
    x_{3,t} & = & 3.4x_{3,t-1}(1 - x_{3,t-1}^2e^{x_{3,t-1}^2}) + c_{13}x_{1,t-1} + 0.4u_{3,t}\nonumber \\
    x_{4,t} & = & 3.4x_{4,t-1}(1 - x_{4,t-1}^2e^{x_{4,t-1}^2}) + c_{54}x_{5,t-3} + c_{64}x_{6,t-1} + 0.4u_{4,t}\nonumber \\
    x_{5,t} & = & 3.4x_{5,t-1}(1 - x_{5,t-1}^2e^{x_{5,t-1}^2}) + 0.4u_{5,t}\nonumber \\
    x_{6,t} & = & 3.4x_{6,t-1}(1 - x_{6,t-1}^2e^{x_{6,t-1}^2}) + c_{76}x_{7,t-3} + 0.4u_{6,t}\nonumber \\
    x_{7,t} & = & 3.4x_{7,t-1}(1 - x_{7,t-1}^2e^{x_{7,t-1}^2}) +  0.4u_{7,t}\nonumber \\
    x_{8,t} & = & 3.4x_{8,t-1}(1 - x_{8,t-1}^2e^{x_{8,t-1}^2}) + c_{78}x_{7,t-1} + 0.4u_{8,t}\nonumber \\
    x_{9,t} & = & 3.4x_{9,t-1}(1 - x_{9,t-1}^2e^{x_{9,t-1}^2}) + c_{79}x_{7,t-1} + 0.4u_{9,t}\nonumber \\
    \label{MVAR}
    \end{eqnarray}
\noindent
with $c_{13}=0.25$, $c_{21}=2.5$, $c_{31}=1.8$, $c_{41}=1.5$, $c_{54}=1.5$, $c_{64}=1.2$, $c_{76}=1.5$, $c_{79}=1.8$, $c_{78}=0.8$ and $u_{n,t}$ being zero mean Gaussian noise. The causal structure of the system described above is shown in Figure \ref{stochastic}. 

We vary the threshold $\delta$ from $0$ to $0.5$, and for each value of $\delta$, we generate 50 realizations for the system described in Equation \eqref{MVAR}. We varied the range of delays from $1$ to $10$ and the embedding dimension and delay were set to $3$ and $100$, respectively. We performed the simulations for generating data sets of a size of $T = 1000$, $5000$, $10000$ and $20000$ samples.

\begin{figure}
    \centering
    \includegraphics[scale=0.35]{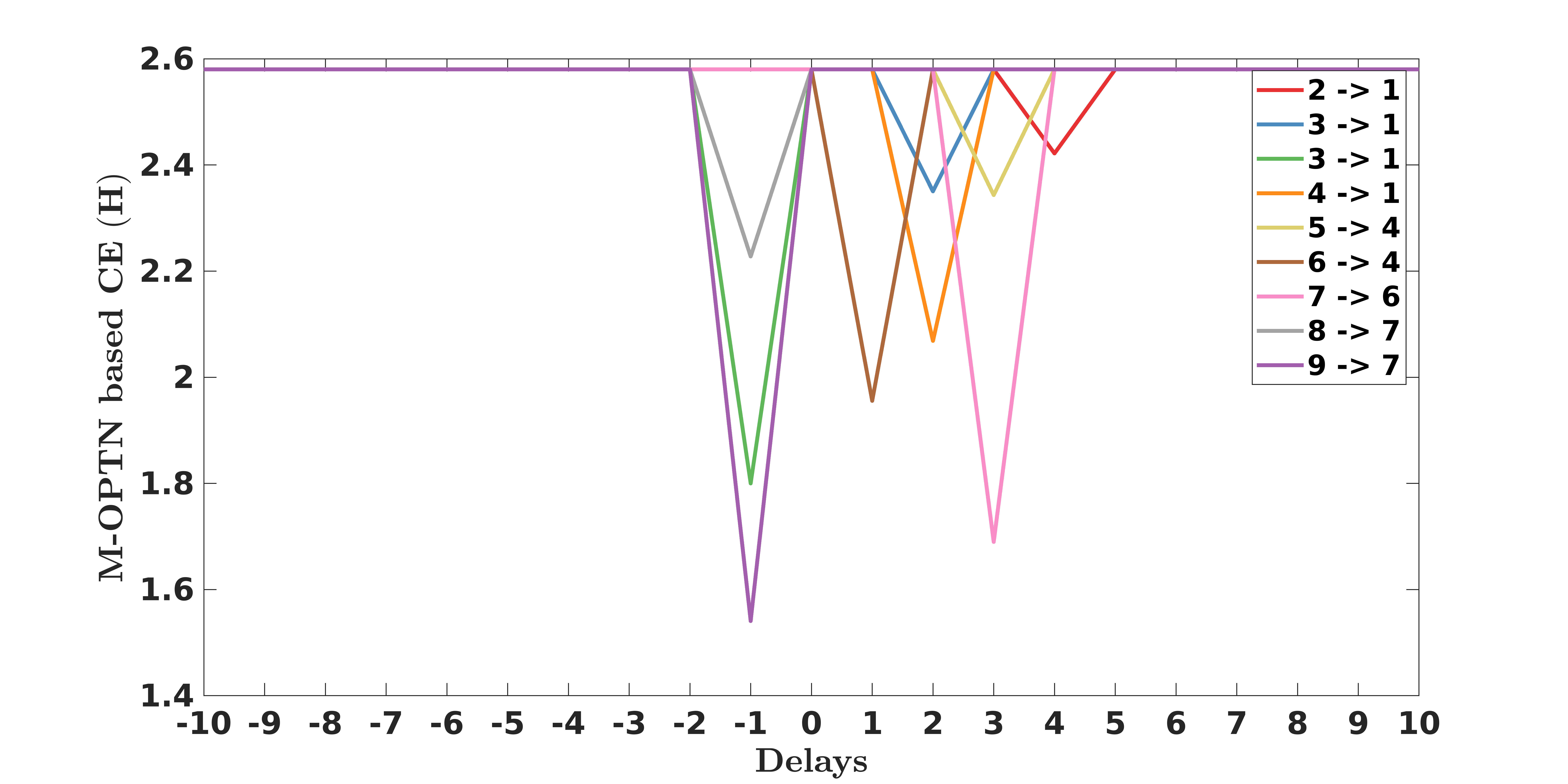}
    \caption{Causality detection based on M-OPTNs for an exemplary simulation of the multivariate autoregressive system for $\delta = 0.15$ and $T=10000$ samples.}
    \label{example_res}
\end{figure}

Figure~\ref{example_res} shows the results from an exemplary simulation, where $\delta = 0.15$, $N=10000$ samples and $NL=0$. As we can see from the figure, the causal interactions among the stochastic processes are correctly identified along with their respective delays. In Figure~\ref{example_res} the interaction $m \rightarrow n$ at a negative delay is again to be interpreted as $n \rightarrow m$. The causal interaction between two processes at a particular delay results in a drop in the CE value away from the $H_{max}$ value, which in our case is given by $\log M! \approx 2.58$. 

\begin{figure}[h]
    \centering
    \includegraphics[scale=0.2]{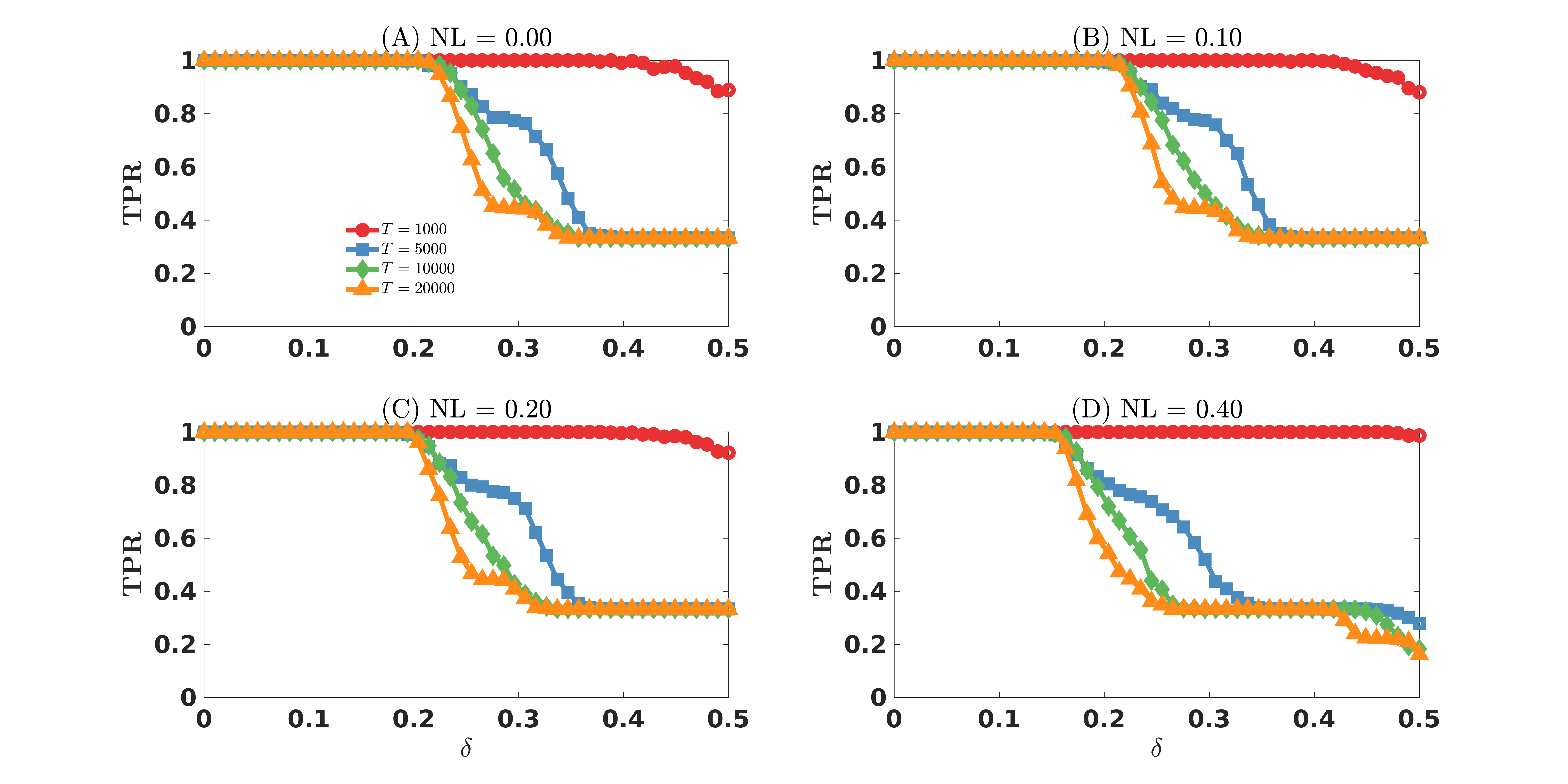}
    \caption{$TPR$ computed for 50 independent realizations of the multivariate autoregressive system as $\delta$, $T$, and $NL$ are varied.}
    \label{tpr_ar}
\end{figure}

\begin{figure}[h]
    \centering
    \includegraphics[scale=0.2]{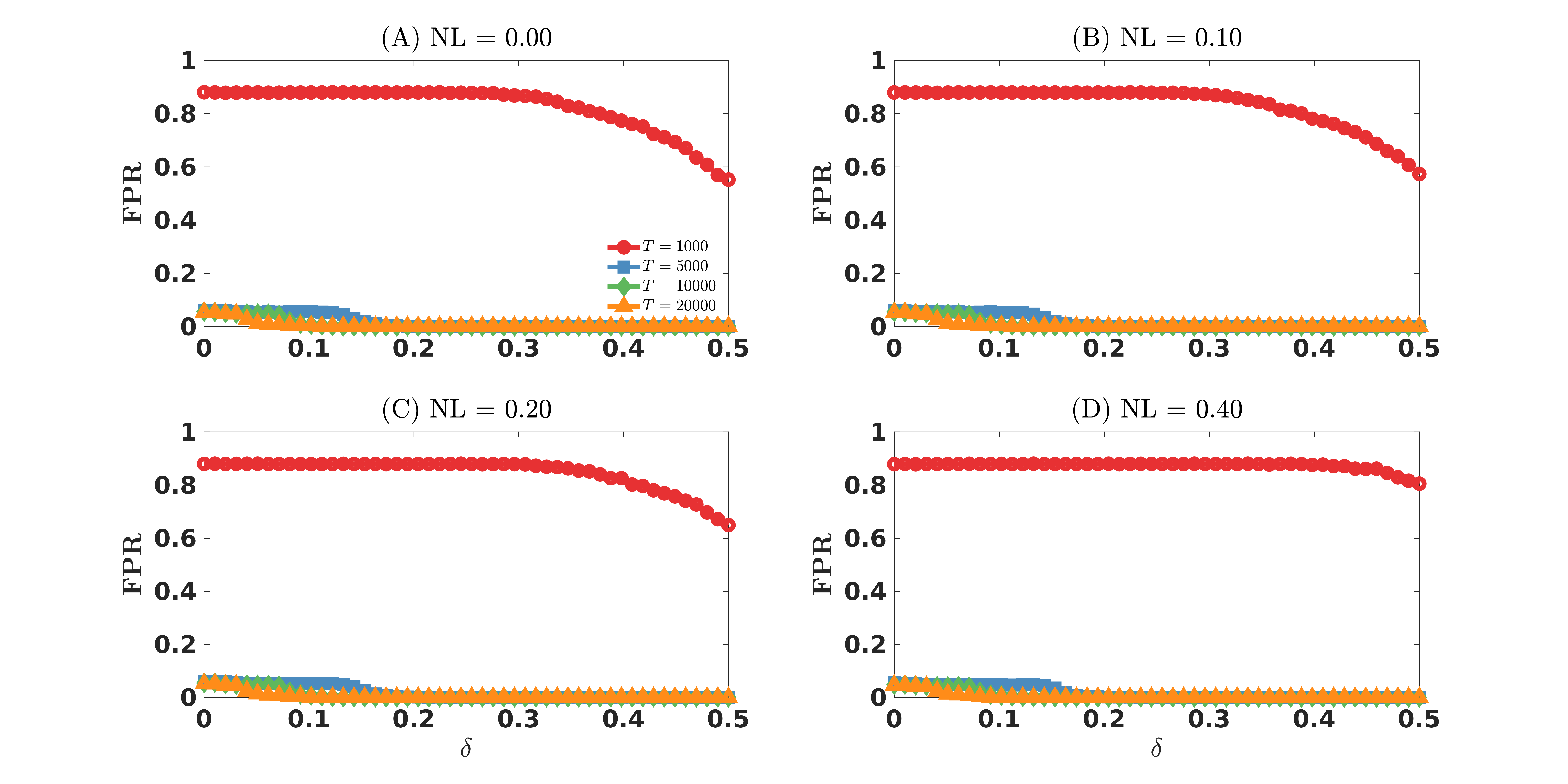}
    \caption{$FPR$ computed for 50 independent realizations of the multivariate autoregressive system as $\delta$, $T$, and $NL$ are varied.}
    \label{fpr_ar}
\end{figure}

Figures~\ref{tpr_ar} and \ref{fpr_ar} show the $TPR$ and $FPR$ values, respectively, for the interacting stochastic system, as $\delta$, $T$ and $NL$ are varied. We can see that at $T=1000$, high $TPR$ as well as $FPR$ values are obtained, with no significant changes in their values as $NL$ or $\delta$ is varied. In case of the $TPR$ values (see Figure~\ref{tpr_ar}), we find that for $T\geq 5000$, the $\delta$ value at which $TPR$ starts to drop below $1$ slightly decreases as $NL$ increases, but in all cases, $TPR=1$ for $\delta<0.2$. For $\delta>0.3$, the $TPR$ values tend to remain at $\approx 0.25$, except for $NL=0.40$, where it continues to decrease for $\delta > 0.4$. When the $T\geq 5000$, we can see that the $FPR$ values drop below $0.1$ and for $\delta>0.2$, $FPR=0$ for $T=5000$, and for $\delta>0.05$ $FPR=0$ for $T=10000$ or $T=20000$, irrespective of the value of $NL$ (see Figure~\ref{fpr_ar}). 

\begin{figure}
    \centering
    \includegraphics[scale=0.2]{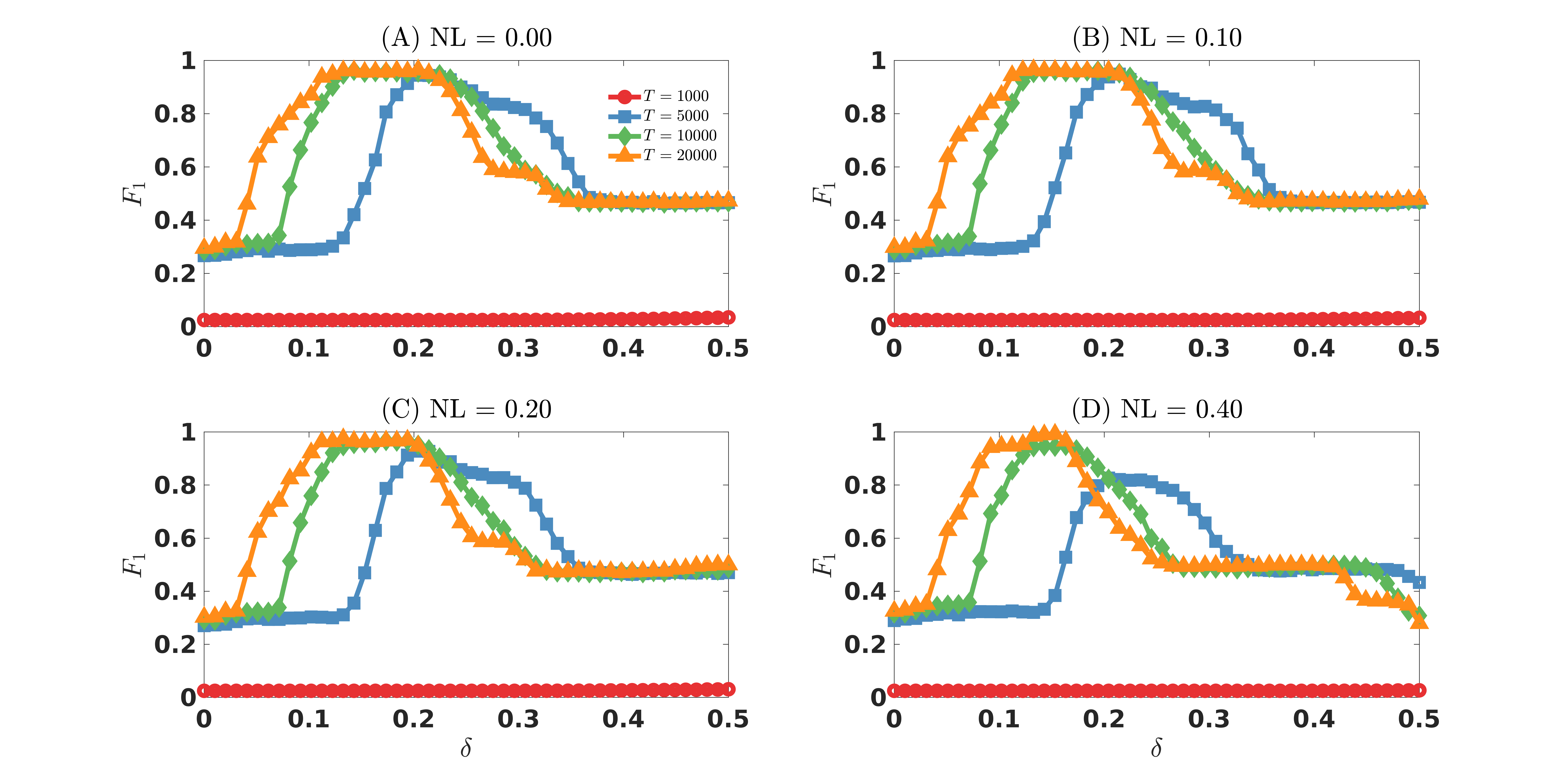}
    \caption{$F_1$ computed for 50 independent realizations of the multivariate autoregressive system as $\delta$, $T$, and $NL$ are varied.}
    \label{f1_ar}
\end{figure}

Figure~\ref{f1_ar} shows the accuracy of the causal inference algorithm in terms of the $F_1$-score for the interacting stochastic system. We observe that for $T=1000$, $F_1=0$, irrespective of the choice of $\delta$ or the level of the noise. For $T=5000$, we find that $F_1$ tends to increase as $\delta$ is increased and reaches $\approx 0.9$ only when $\delta >0.2$. $F_1$ tends to decrease for $\delta > 0.25$ for $NL < 0.40$. At $NL=0.40$, for $T=5000$, $F_1$ reaches a maximum value of $0.8$ for $0.2 < \delta < 0.25$, after which it tends to decrease. For $T=10000$ or $20000$, we again observe that $F_1$ tends to increase when $\delta$ is increased and it reaches $\approx 1$ for the range $0.1 < \delta < 0.2 $. The range of $\delta$ tends to get narrower only at $NL=0.40$, when $F_1 \approx 1$ for $0.1 < \delta < 0.15$. $F_1$ starts to decrease as $\delta > 0.2$ for $NL < 0.4$ and $\delta > 0.15$ for $NL=0.4$.  

\subsection{Interacting Lorenz systems}\label{sec:3.2}
The next example of three interacting identical Lorenz systems with the structure $X_1 \rightarrow X_2 \rightarrow X_3 $ is defined by the following set of ordinary differential equations,
\begin{align}
    \frac{dx_1(t)}{dt} &= 10(y_1(t) - x_1(t)) \nonumber \\
    \frac{dy_1(t)}{dt} &= 28x_1(t) - y_1(t) - x_1(t)z_1(t) \nonumber \\
    \frac{dz_1(t)}{dt} &= x_1(t)y_1(t) - 8/3z_1(t) \nonumber \\
    \frac{dx_2(t)}{dt} &= 10(y_2(t) - x_2(t)) + c(x_1(t) - x_2(t)) \nonumber \\
    \frac{dy_2(t)}{dt} &= 28x_2(t) - y_2(t) - x_2(t)z_2(t) \nonumber \\
    \frac{dz_2(t)}{dt} &= x_2(t)y_2(t) - 8/3z_2(t) \nonumber \\
    \frac{dx_3(t)}{dt} &= 10(y_3(t) - x_3(t)) + c(x_2(t) - x_3(t)) \nonumber \\
    \frac{dy_3(t)}{dt} &= 28x_3(t) - y_3(t) - x_3(t)z_3(t) \nonumber \\
    \frac{dz_3(t)}{dt} &= x_3(t)y_3(t) - 8/3z_3(t) 
\end{align}
We have set the coupling strength to $c=0.6$ and used a Runge-Kutta integrator with a step size of $dt=0.001$ to numerically solve the above set of ordinary differential equations. We then used the time series $\{x_k(t)\}_{t=1}^T$ with $k=1, 2, 3$ as the observations from the interacting Lorenz systems, to which we added noise at $NL = 0.1, 0.2, 0.4$. We set $\lambda=0.995$ and performed 50 simulations for every combination of the threshold $\delta$, time series length $T$ and noise level $NL$ including the noise-free condition (i.e., $NL=0$).

\begin{figure}
    \centering
    \includegraphics[scale=0.2]{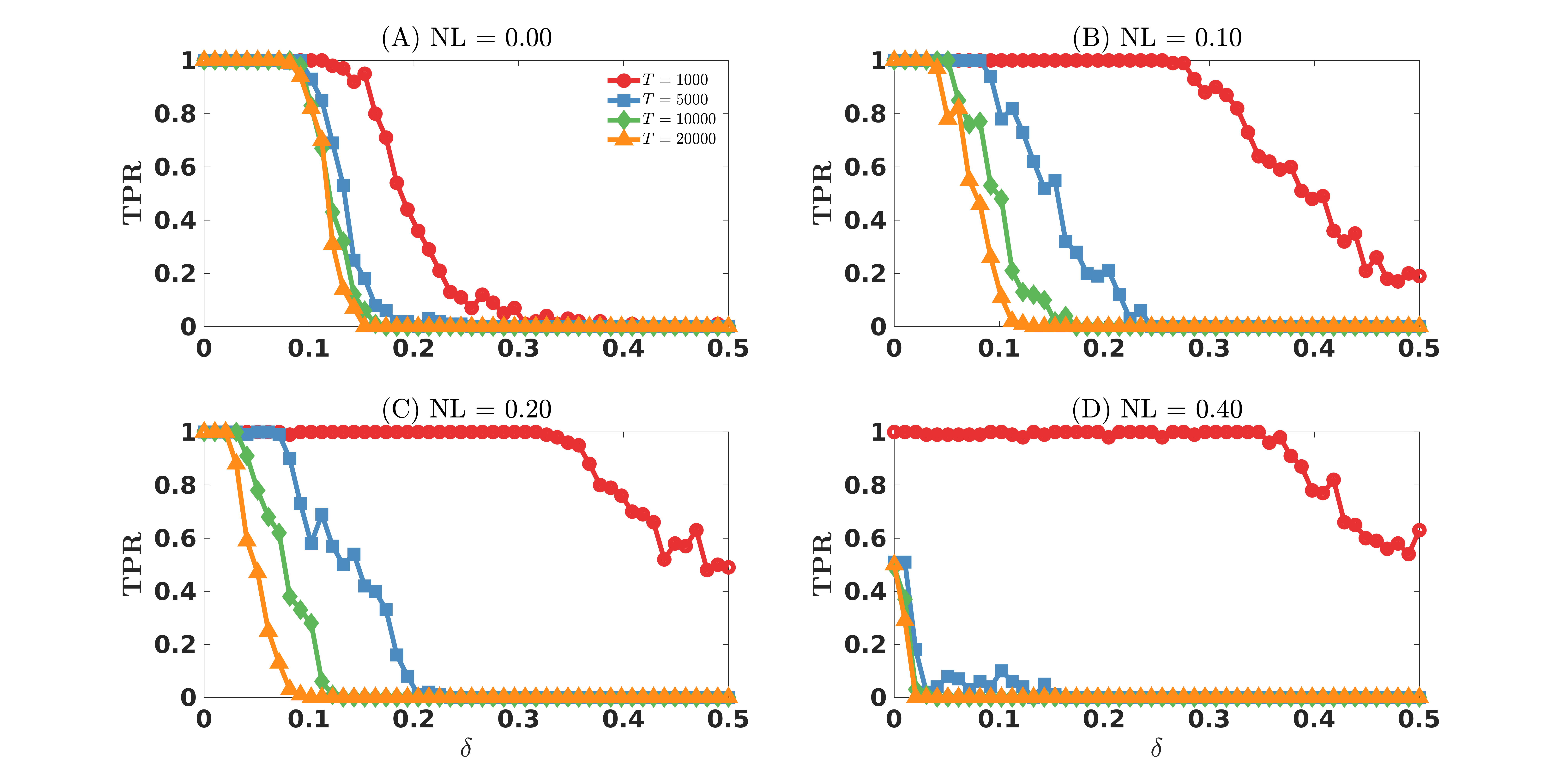}
   \caption{TPR of the proposed causal inference algorithm for the interacting Lorenz systems as $\delta$, $T$, and $NL$ are varied.}
    \label{tpr_lorenz}
\end{figure}

\begin{figure}
    \centering
    \includegraphics[scale=0.2]{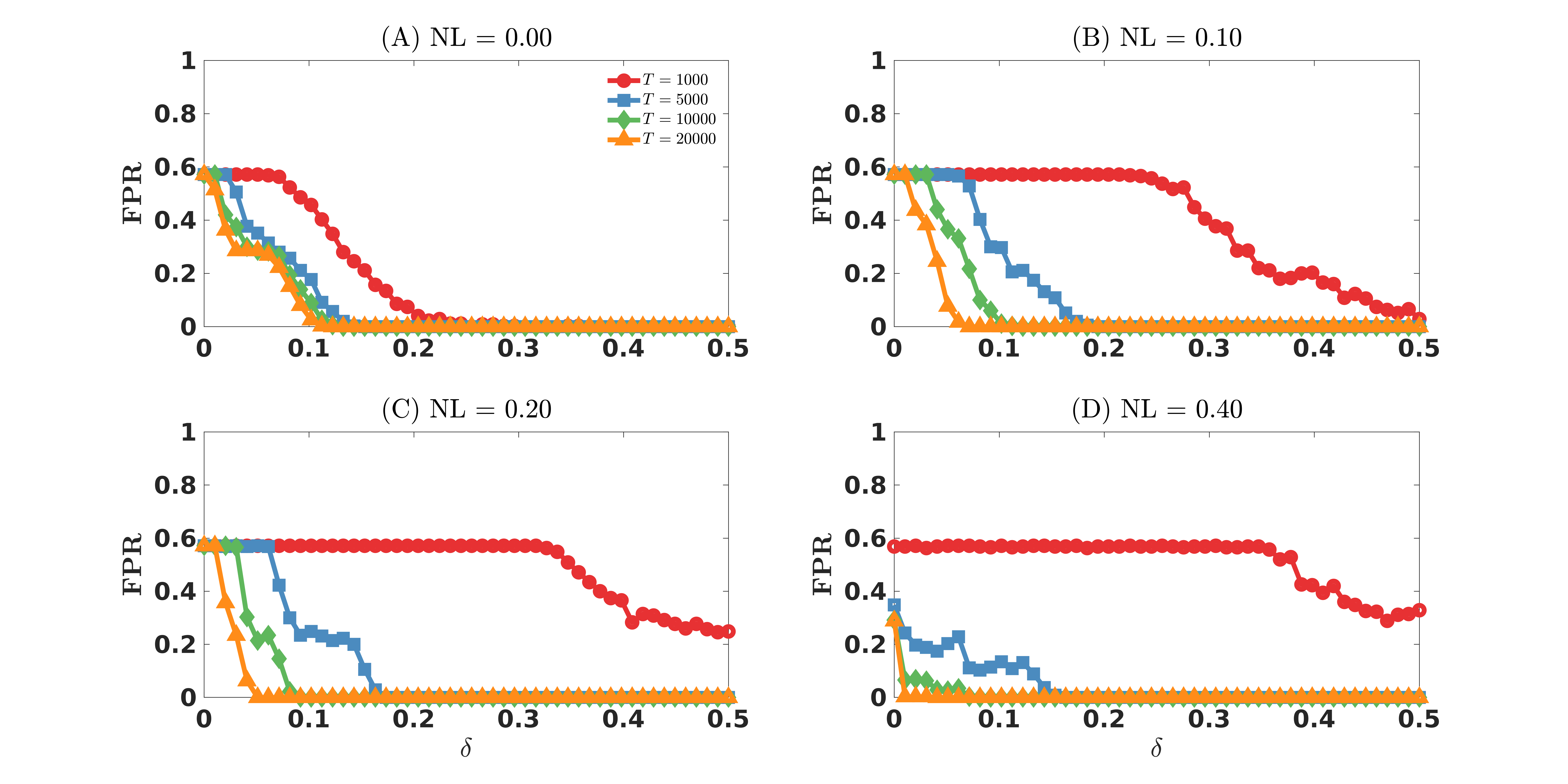}
   \caption{FPR of the proposed causal inference algorithm for the interacting Lorenz systems as $\delta$, $T$, and $NL$ are varied.}
    \label{fpr_lorenz}
\end{figure}

\begin{figure}
    \centering
    \includegraphics[scale=0.2]{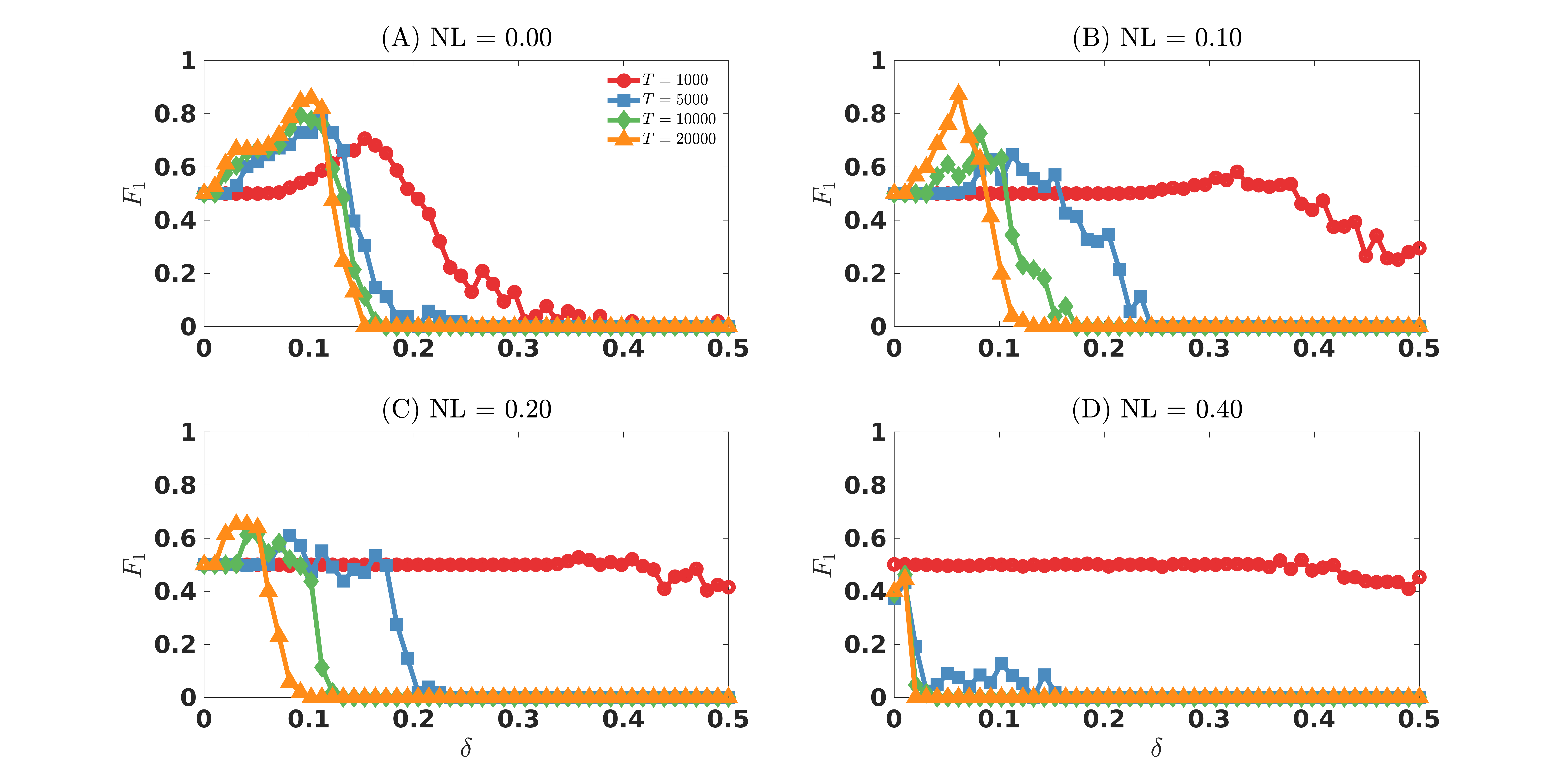}
   \caption{$F_1$-score of the proposed causal inference algorithm for the interacting Lorenzs system as $\delta$, $T$, and $NL$ are varied.}
    \label{f1_lorenz}
\end{figure}

Figures~\ref{tpr_lorenz} and \ref{fpr_lorenz} show the $TPR$ and $FPR$ values obtained for the interacting Lorenz systems as $\delta$ and $T$ is varied for different values of $NL$. Here $NL = 0$ represents the noise-free condition. We can see that for $T=1000$, the proposed causal inference algorithm gives high $TPR$ ($\approx 1$) as well as $FPR$ ($\approx 0.60$), which starts to drop for $\delta>0.15$ and $\delta>0.1$, respectively, for $NL = 0$. As $NL$ is increased, the $\delta$ at which $TPR$ and $FPR$ start to drop, also increases. However, we observe that for all values of $NL$, for $T=1000$, the accuracy of the algorithm as given by the $F_1$ score (see Figure~\ref{f1_lorenz}) is the highest at $\delta \approx 0.2$ for $NL=0$ and for other values of $NL$, $F_1$ remains mostly at $0.5$. 

For $T=5000$, the $TPR$ remains at $1$ for $NL = 0$, $0.10$ and $0.20$ and starts to drop for $\delta > 0.1$ (see Figure ~\ref{tpr_lorenz}). Only in the case of $NL=0$, the $FPR$ drops to $\approx 0.20$ at $\delta \approx 0.08$ (see Figure ~\ref{tpr_lorenz}). For $NL = 0.10$ and $0.20$, FPR remains at $0.6$ and only starts to drop when $\delta > 0.1$. For $NL=0.40$, we observe low $FPR$ ($< 0.4$) as well as $TPR$ ($<0.5$) values for all values of $\delta$. Figure~\ref{f1_lorenz} shows that for $T=5000$, at $NL=0.0$ and $\delta \approx 0.15$, the $F_1$-score is $\approx 0.78$ and starts to drop to 0 and $\delta$ is increased. The  $F_1$-score remains mostly at about $0.5$ and drops to $0$ for other values $NL$ as $\delta$ is increased. 

From Figure~\ref{tpr_lorenz}, for $T=10000$ and $20000$, we observe that $TPR$ remains at $1$ and the $\delta$ value at which it starts to drop and eventually reaches zero decreases as $NL$ increases. However, we find $TPR > 0.8$ even with the addition of noise at $NL=0.1$ and $0.2$, for $\delta < 0.05$, for which we also observe that the $FPR$ starts to drop (see Figure~\ref{fpr_lorenz}). This is also reflected in the accuracy as given by the $F_1$-score shown in Figure~\ref{f1_lorenz}, where $F_1 \approx 0.88$ for $NL=0$ and $\delta=0.1$, and  $F_1 \approx 0.88$ for $NL=0.1$ and $\delta \approx 0.05$. As $NL$ is increased to $0.20$, $F_1$-score drops to $0.6$. At $NL=0.40$, we observe that the accuracy is $0$ for all values of $\delta$, except for very small values ($<0.03$), for which $F_1 \approx 0.4$. 

\subsection{Network of neural mass models}\label{sec:3.3}
The simulations described in Section~\ref{sec:2} and \ref{sec:3.2} have been restricted to linear stochastic systems and paradigmatic nonlinear dynamical systems, both of which may not fully characterize the typical nonlinear characteristics in neural time series. In order to also demonstrate the ability of the proposed method to capture interactions in nonlinear dynamical systems such as neuronal networks, we finally consider a network of neural mass models \cite{Ref22}. To this end, we created a network of eight neural mass models (the ordinary differential equations describing each neural mass model are provided in the Appendix, and the parameters are set as given in \cite{Ref22}), with the $\%$ of directed interactions ($\mathcal{K}$) between the eight regions varying as  $5\%$, $10\%$ and $25\%$ of the overall possible connections ($N^2$ including $N$ self-connections), at a delay of $40$ milliseconds. We vary the threshold $\lambda$ from $0.99$ to $1.0$ and the threshold $\delta$ from $0$ to $0.2$. To also investigate the effect of noise on the performance of the method -- in addition to the noise free observations from neural mass models, Gaussian noise at a $NL$ of $0.5$ and $1.0$ was added to the output of the neural mass models. For each of the combinations of these parameters ($\mathcal{K}$, $\lambda$, $\delta$, and $NL$), we generated $25$ simulations. The embedding parameters were the same as in Section~\ref{sec:2} ($M=3$ and $d=1$). We computed CE based on M-OPTNs for delays ranging from $10$ milliseconds to $100$ milliseconds. Since the interaction in the simulated network happens at around $40$ milliseconds, any interaction in the estimated networks at a delay other than $40$ milliseconds is counted as a false positive. 

\begin{figure}[h]
    \centering
    \includegraphics[scale=0.3]{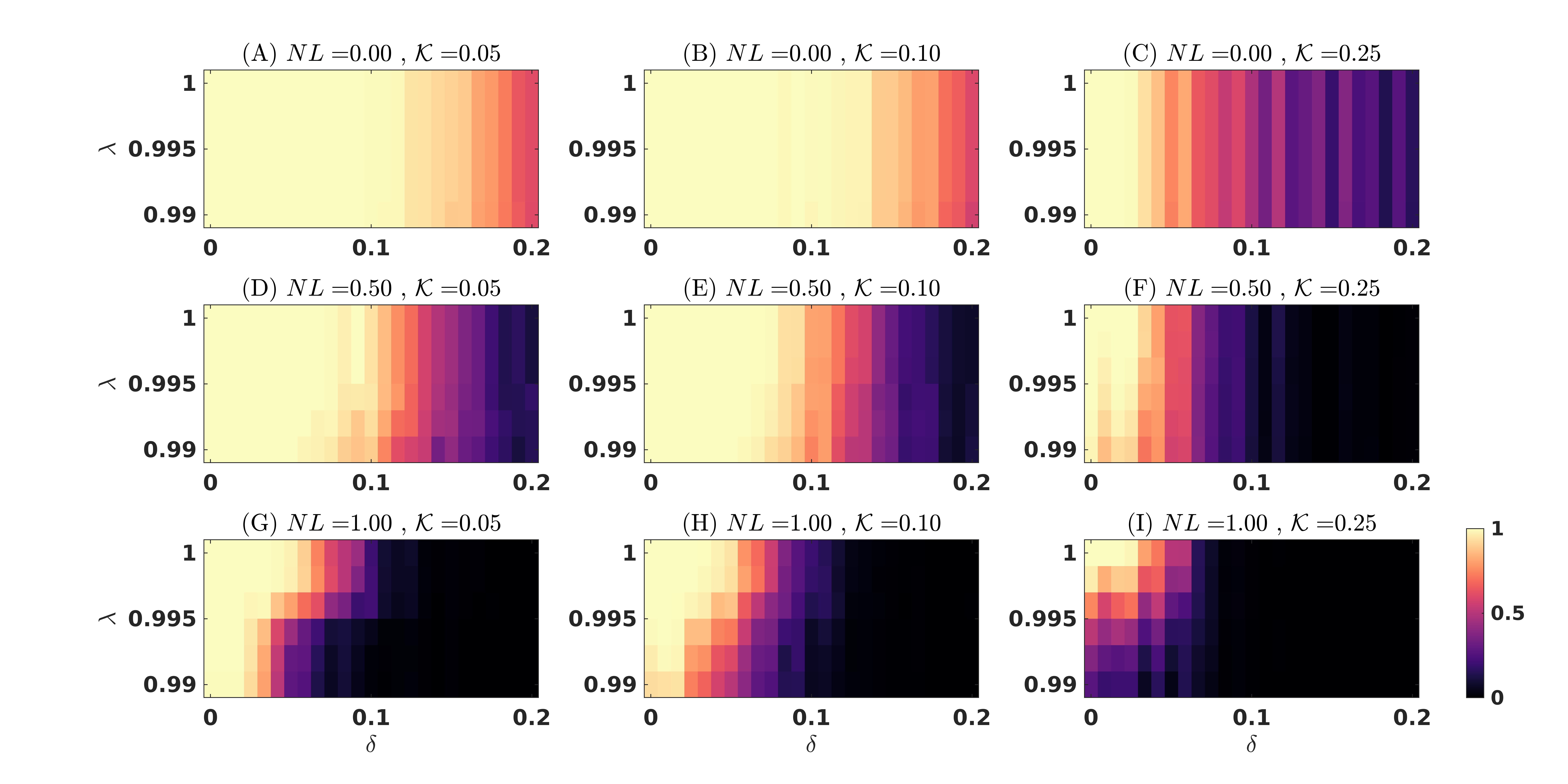}
    \caption{TPR computed for 50 independent realizations of networks of neural mass models for every combination of $\mathcal{K}$, $\lambda$, $\delta$, and $NL$.}
    \label{tpr_NMM}
\end{figure}

Figure~\ref{tpr_NMM} shows the $TPR$ values for the network of neural mass models for varying $NL$, $\lambda$, $\delta$ and $\mathcal{K}$. We can see that for all values of $\mathcal{K}$ and $NL=0$ and $0.5$, the value of $\delta$ at which $TPR$ remains at $1$ decreases and $TPR$ values are largely unaffected by the choice of $\lambda$. At $NL=1.0$, we observe that as $\lambda$ is increased, the range of $\delta$ for which $TPR=1$ increases and at higher connection densities ($\mathcal{K}=0.25$), we observe that high $TPR$ values are only observed for a narrow range of $\lambda>0.995$ and $\delta$ values. 

\begin{figure}[h]
    \centering
    \includegraphics[scale=0.3]{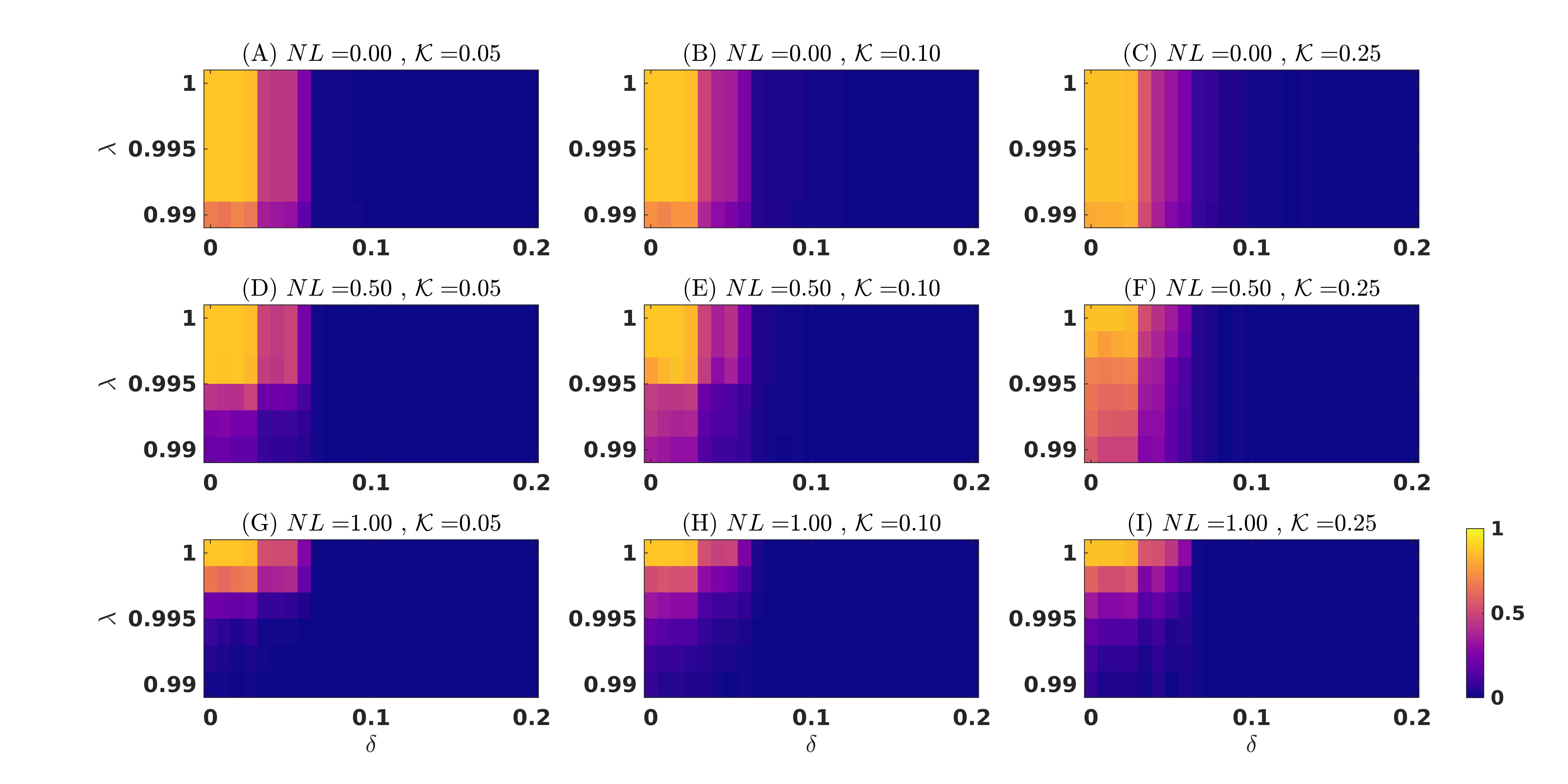}
    \caption{FPR computed for 50 independent realizations of networks of neural mass models for every combination of $\mathcal{K}$, $\lambda$, $\delta$, and $NL$.}
    \label{fpr_NMM}
\end{figure}

Figure~\ref{fpr_NMM} shows the associated $FPR$ values. We observe that for $\delta < 0.03$, $FPR$ remains very high and close to $1$ irrespective of the choice of $\delta$, $\lambda$  or $\mathcal{K}$ when the $NL = 0$ or $0.5$. When $NL=1.0$, high values of $FPR$ are observed for $\lambda>0.998$. When $\delta>0.05$ and $NL=0$ or $0.5$, the $FPR$ values drop to zero irrespective of the choice of $\lambda$ and for all $\mathcal{K}$. When $NL$ is increased to $1.0$, we observe in general lower values of $FPR$ as $\lambda$ is decreased. However for $\delta>0.05$, $FPR$ drops to zero irrespective of the choice of $\lambda$ or $\mathcal{K}$.

\begin{figure}[h]
    \centering
    \includegraphics[scale=0.3]{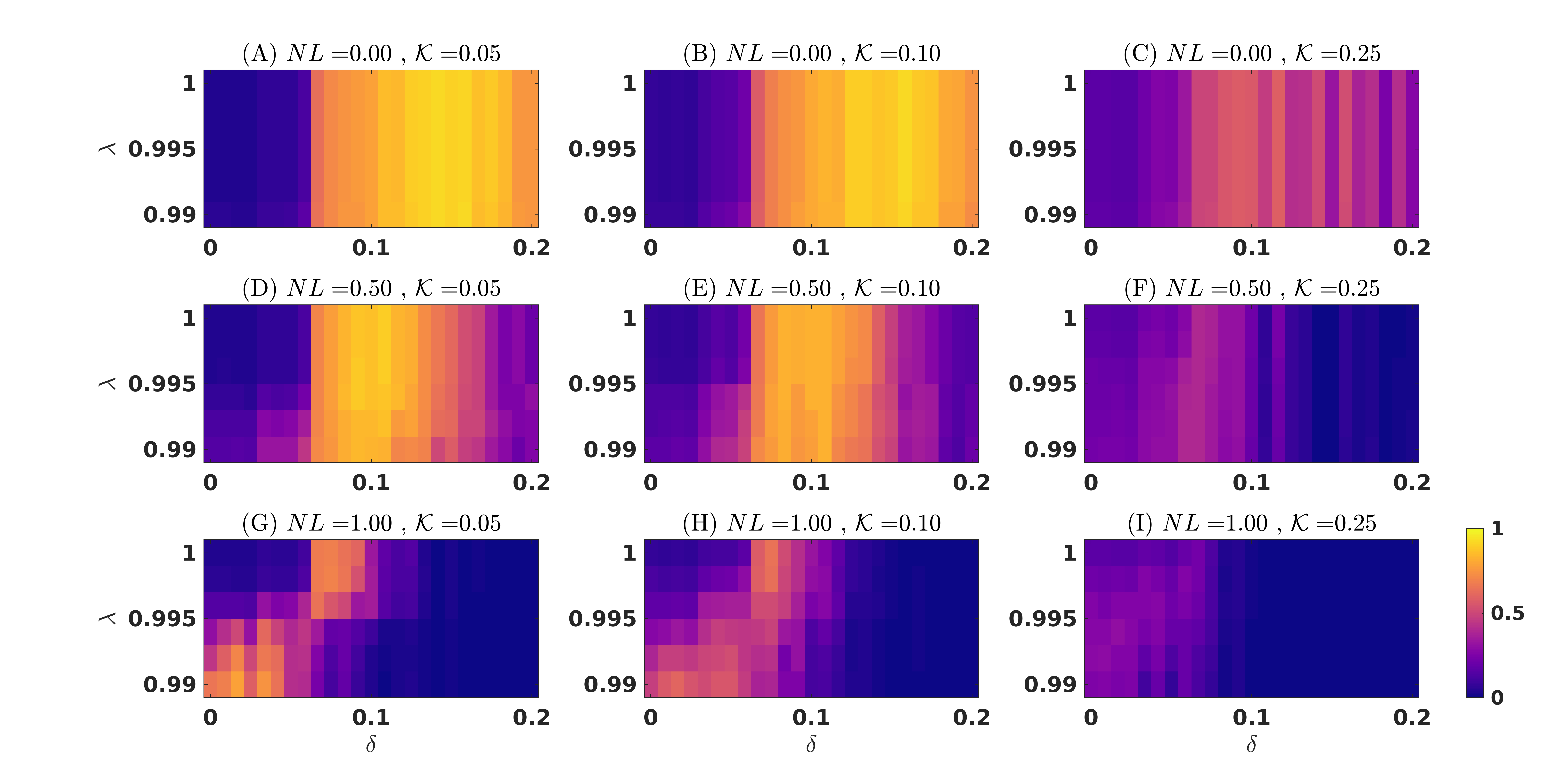}
    \caption{$F_1$-score computed for 50 independent realizations of networks of neural mass models for every combination of $\mathcal{K}$, $\lambda$, $\delta$, and $NL$.}
    \label{f1_NMM}
\end{figure}

The accuracy of the proposed approach for the network of neural mass models is finally shown in terms of the $F_1$-score in Figure~\ref{f1_NMM}. As it is evident from the figure, as $NL$ and $\mathcal{K}$ increases, the choice of $\lambda$ and $\delta$ for which we obtain high values for $F_1$ gets narrower. For example, at $NL=0$ and $\mathcal{K}=0.1$, for $0.1 < \delta < 0.2$, we can see that $F_1 \approx 1$ irrespective of the choice of $\lambda$. But for the same $\mathcal{K}$, we observe that in order to get good accuracy, i.e., high $F_1$-score,  we need to choose lower values of $\lambda$ and $\delta$ as $NL$ is increased. When $\mathcal{K} = 0.25$, and $NL <= 0.5$, we observe that $F_1 \approx 0.7$ for certain choices of $\lambda$ and $\delta$ but when $NL=1.0$, $F_1<0.5$.

\section{Causality detection in MEA electrophysiology data}
To validate our approach against real-world experimental data, we have applied the developed algorithm to electrophysiological recordings of epileptiform patterns generated by 4-aminopyridine (4AP)-treated rodent hippocampus-cortex (CTX) slices. In order to visualize the network activity, we transform the CE values obtained from two MEA signals $i$ and $j$, $H(i \vert j)$ as
\begin{equation}
    S(i,j) = 2.58 - H(i,j)
    \label{norm}
\end{equation}
and normalize them such that the strongest pairwise interaction takes a value of $1$, i.e., $\tilde{S} = S(i,j)/\max(S)$.


\subsection{Brain slice preparation and maintenance} 
Combined hippocampus-cortex (CTX) brain slices ($n=4$), $400~\mu$m thick, were prepared from four male CD1 mice (4-8 weeks old), as previously described \cite{Ref25}. Briefly, animals were euthanized under deep isoflurane anesthesia, their brain was quickly removed and placed into ice-cold ($\approx 2\degree$C) sucrose-based artificial cerebrospinal fluid (sucrose-ACSF) composed of (mM): Sucrose 208, KCl$_2$, KH$_2$PO$_4$ 1.25, MgCl$_2$ 5, MgSO$_4$, CaCl$_2$ 0.5, D-glucose 10, NaHCO$_3$ 26, L-Ascorbic Acid 1, Pyruvic Acid 3. The brain was let to chill for 2~min before slicing in ice-cold sucrose-ACSF using a vibratome (Leica VT1000S, Leica, Germany). Brain slices were immediately transferred to a submerged holding chamber containing room-temperature holding ACSF composed of (mM): NaCl 115, KCl$_2$, KH$_2$PO$_4$, 1.25, MgSO$_4$ 1.3, CaCl$_2$ 2, D-glucose 25, NaHCO$_3$ 26, L-Ascorbic Acid 1. After at least 60 minutes recovery, individual slices were transferred to a submerged incubating chamber containing warm ($\approx 32\degree$C) holding ACSF for 20-30 minutes and subsequently incubated in warm ACSF containing the K+ channel blocker 4-aminopyridine (4AP, $250$ $\mu M$), in which MgSO$_4$ concentration was lowered to 1 mM (4AP-ACSF, \cite{Ref25}). Brain slice treatment with 4AP is known to enhance both excitatory and inhibitory neurotransmission and induces the acute generation of epileptiform discharges \cite{Ref31}. All brain slices were incubated in 4AP-ACSF for at least 1 hour before beginning any recording session. All solutions were constantly equilibrated at pH$\approx 7.35$ with $95\%$ O$_2$ / $5\%$ CO$_2$ gas mixture (carbogen) and had an osmolality of 300-305 mOsm/kg. Chemicals were acquired from Sigma-Aldrich. 
All procedures have been approved by the Institutional Animal Welfare Body and by the Italian Ministry of Health (authorization 176AA.NTN9), in accordance with the National Legislation (D.Lgs. 26/2014) and the European Directive 2010/63/EU. All efforts were made to minimize the number of animals used and their suffering.

\subsubsection{MEA recording and signal pre-processing} 
Individual brain slices were placed on a $6\times10$ planar MEA (TiN electrodes, diameter 30~$\mu$m, inter-electrode distance 500~$\mu$m, impedance $<$ 100 k$\Omega$), held in place by a custom-made anchor, and continuously perfused at $\approx 1$ ml/min with 4AP-ACSF at ($\approx 32\degree$C), equilibrated with carbogen gas mixture. To allow for laminar flow and a high exchange rate of the 4AP-ACSF, a custom-made low-volume ($\approx 500\mu$l) recording chamber (Crisel Instruments, Italy) replaced the default MEA ring \cite{Ref25}.

Extracellular field potentials were acquired at 5 kHz (pre-sampling low-pass filter at 2 kHz) using the MEA2100-mini-HS60 system through the Multichannel Experimenter software (all from Multichannel Systems – MCS, Reutlingen, Germany) and stored on the hard drive for off-line analysis.

For the implementation of the M-OPTN, signals were pre-processed by low-pass filtering ($1$ kHz) and resampling at $3$ kHz. The signals were then further divided into overlapping ($50\%$), $4$ s windows, to obtain time-varying measures of causality.

\subsubsection{Epileptiform activity generated by 4AP-treated hippocampus-CTX slices}
\begin{figure}[htb]
    \centering
    \includegraphics[scale=0.3]{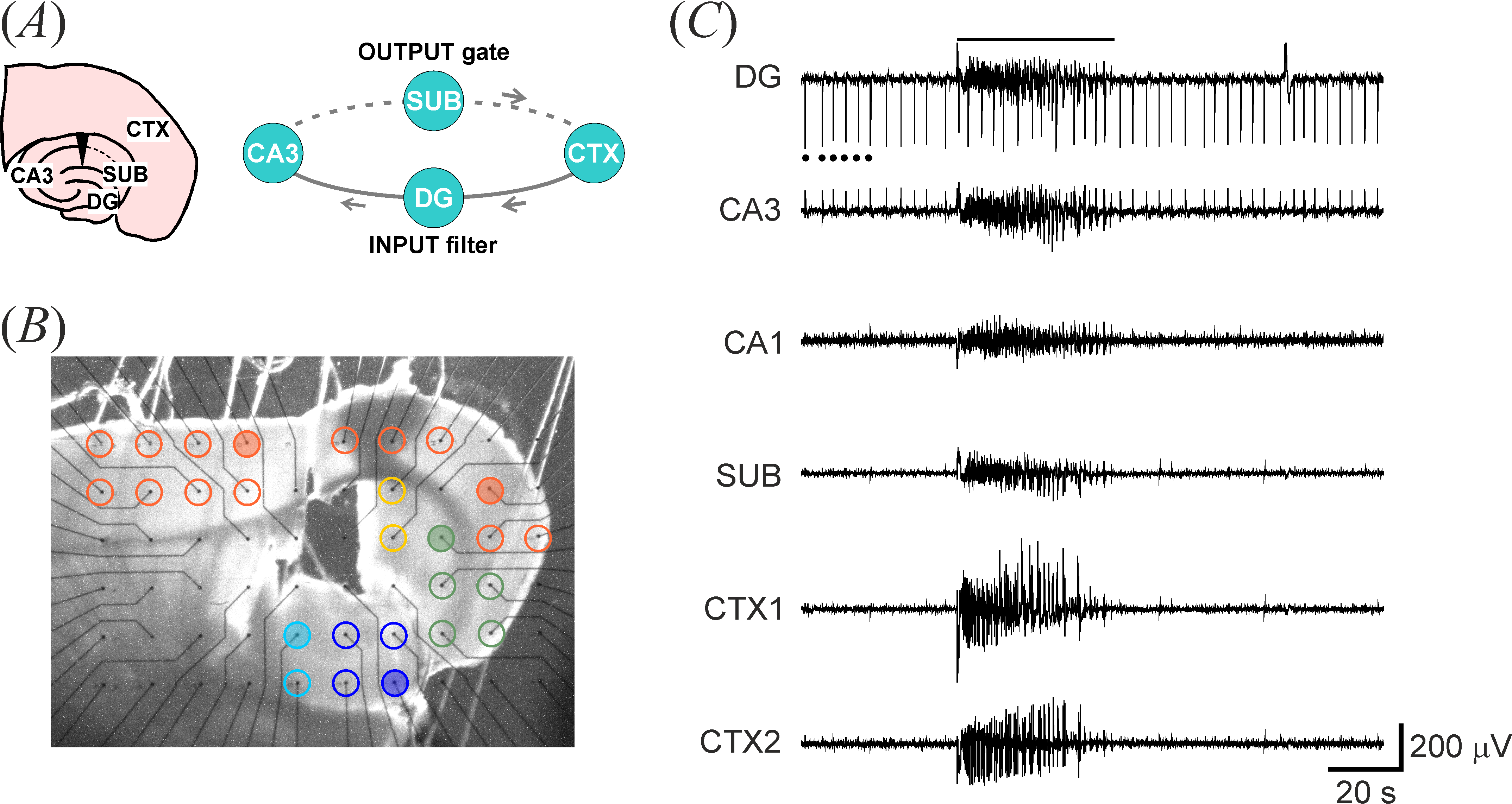}
    \caption{ (A) Schematic rendition of a disconnected hippocampus-CTX slice and its corresponding circuit. The solid line indicates the preserved pathway, whereas the dashed line indicates the disrupted pathway. Arrows indicate the signal propagation along the loop. (B) Hippocampus-\textcolor{black}{CTX} slice laying on a $6 \times 10$ planar MEA. MEA electrodes are placed in DG (blue), SUB (green), CA1 (yellow), CA3 (cyan) and CTX (orange). (C) MEA recording of the epileptiform pattern generated by the brain slice in B. The pattern consists of short interictal events (dots) and prolonged ictal discharges (solid line). Note that the small-amplitude events in CA1, SUB and CTX are far fields originating in CA3.} 
    \label{ex_mea}
\end{figure}

The brain slice preparation used in this work includes the fundamental circuits involved in the generation of limbic seizures seen in temporal lobe epilepsy and enables analyzing the network interactions leading to seizure-like discharge generation. As shown in Figure~\ref{ex_mea}, the key regions of interest (ROIs) in this brain slice preparation are the dentate gyrus (DG), the hippocampal subfields Cornu Ammonis 3 and 1 (CA3 and CA1, respectively), the subiculum (SUB) and the parahippocampal cortex (CTX-1 and CTX-2). These regions communicate through the so-called hippocampal loop \cite{Ref23} (see Figure~\ref{ex_mea} (A)). When challenged with convulsant drugs, such as 4AP, hippocampus-CTX slices generate a typical epileptiform pattern made of three types of activity \cite{Ref24}: (i) slow interictal events, recurring at 5-20 s interval, generated by and spreading to any ROI with no specific site of origin, (ii) fast interictal events, recurring at 0.5-2 s interval, generated specifically by the CA3, propagating to the CA1 via the Schaffer Collaterals and subsequently reaching the CTX through the SUB (output gate), (iii) ictal (seizure-like) discharges, recurring at 3-5 min interval, originating primarily in the CTX and spreading to the hippocampus proper via the DG (input filter). It has been previously demonstrated that when the hippocampal loop circuitry is intact (connected brain slice), the fast CA3-driven interictal activity controls ictal discharge generation by the CTX, for which ictal discharges disappear within 1-2 hours of 4AP application, while only the interictal patterns remain. At variance, the disruption of the hippocampal loop upon Schaffer Collaterals damage, as seen in hippocampal sclerosis typical of temporal lobe epilepsy (disconnected brain slice) releases the CTX from the CA3 control permitting ictal activity generation and propagation \cite{Ref21}. 

Here, we have analyzed MEA recordings of epileptiform activity generated by disconnected hippocampus-CTX brain slices, in which the Schaffer Collaterals were mechanically severed.

The circuit diagram of a disconnected hippocampus-CTX slice is depicted in Figure \ref{ex_mea} (A) and Figure~\ref{ex_mea} (B) shows a disconnected hippocampus-CTX slice placed on a $6 \times 10$ planar MEA, while Figure~\ref{ex_mea} (C) shows the typical epileptiform pattern induced by 4AP in this brain slice preparation, consisting of brief interictal events and prolonged ictal discharges.

For the purpose of this study, we have selected the signals from six electrodes in each brain slice, to include each of the four hippocampal regions and two CTX locations, one proximal and one distal to the hippocampus with regards to the signal propagation pathway (see Figure~\ref{ex_mea} (B)). For each MEA recording, we have selected a portion of the signal to include ictal activity preceded and followed by $30-100$ s of interictal activity.

The CE based on M-OPTN was computed for each window and the resulting CE value was assigned to the mid-point of each window, to obtain a time-varying measure. 

\subsection{Results}
\begin{figure}
    \centering
    \includegraphics[scale=0.28]{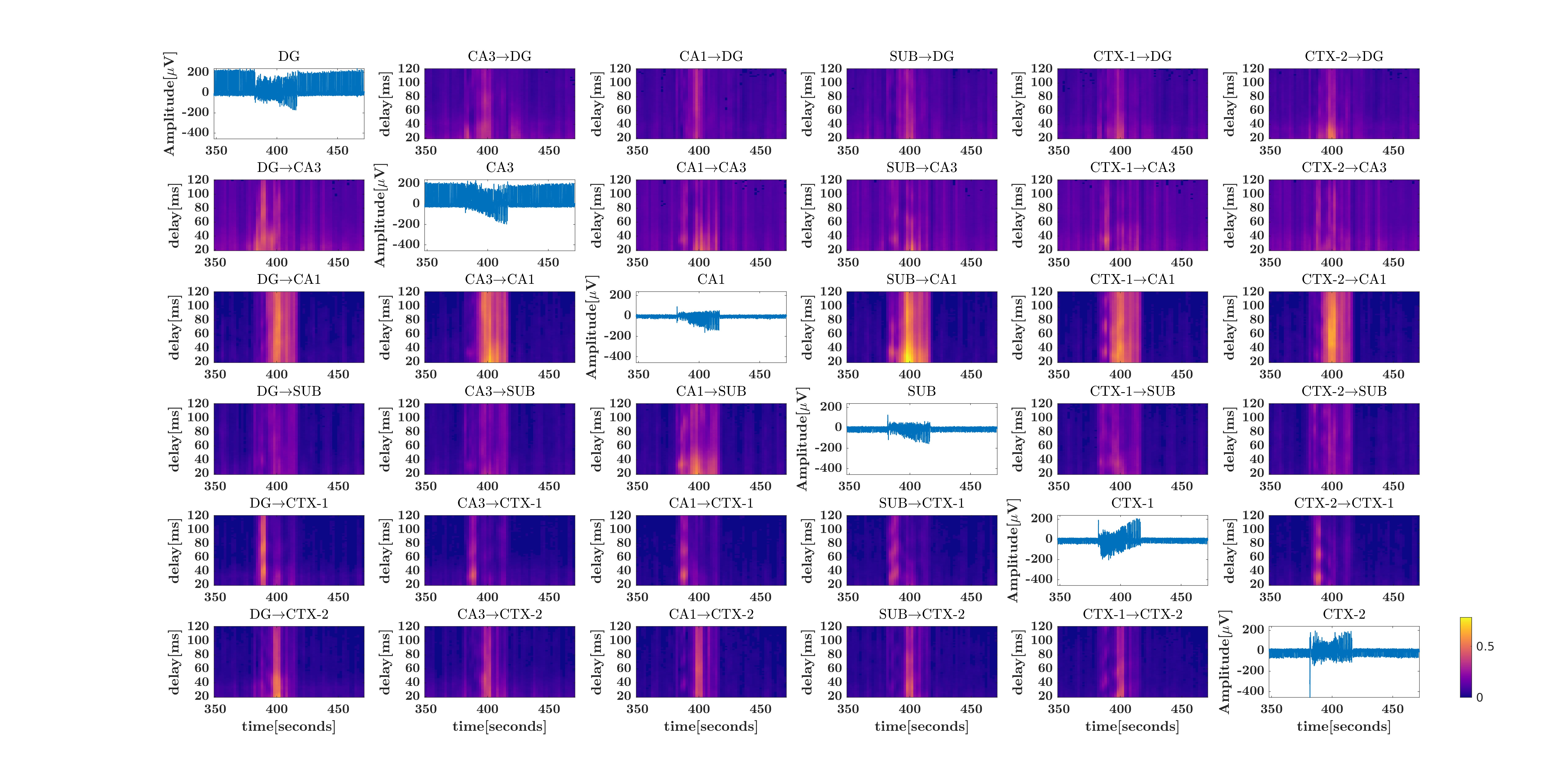}
    \caption{Causality detection based on M-OPTNs for MEA recording of a representative brain slice. The subplots on the diagonal show the MEA signal (ictal activity preceded and followed by interictal events) recorded by the electrodes placed in each of the six ROIs. Each off-diagonal subplot represents the time-varying interaction as given by CE based on M-OPTNs between two ROIs for varying time delays. The color bar indicates the connection strength.}
    \label{mea_1}
\end{figure}

The results from the application of our method are shown in Figure~\ref{mea_1}, where the plots on the main diagonal show the MEA signals acquired from the six selected ROIs. The off-diagonal plots represent the time-varying interaction between two ROIs across varying time delays ($20$ ms to $120$ ms). 
From Figure~\ref{mea_1}, it can be seen that significant network activity starts around the time of the ictal onset, with the strongest connections following the propagation paths CTX $\rightarrow$ DG, CTX $\rightarrow$ CA1, CTX $\rightarrow$ SUB, SUB $\rightarrow$ CTX and DG $\rightarrow$ CA3. The strongest interactions are observed at $38$ and $70$ ms. 

\begin{figure}
    \centering
    \includegraphics[scale=0.28]{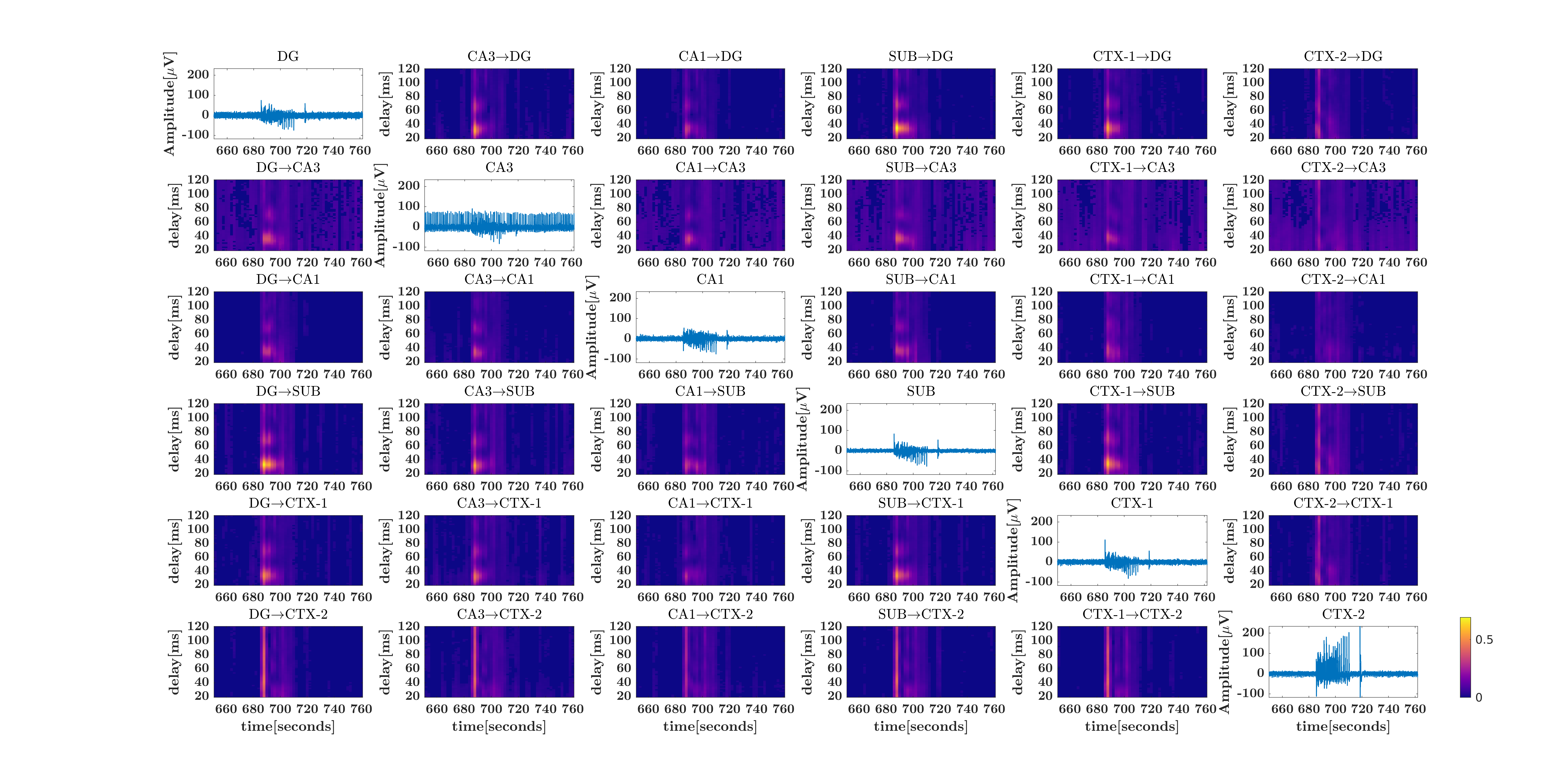}
    \caption{Causality detection based on M-OPTNs for MEA data from slice no. 2. Each off-diagonal subplot represents the time-varying interaction as given by CE based on M-OPTNs between two ROIs for varying time delays. The subplots on the diagonal show the MEA signal (including the ictal activity preceded and followed by inter-ictal activity) in the electrodes placed in each of the six ROIs. The color bar indicates the connection strength.}
    \label{mea_2}
\end{figure}

\begin{figure}
    \centering
    \includegraphics[scale=0.28]{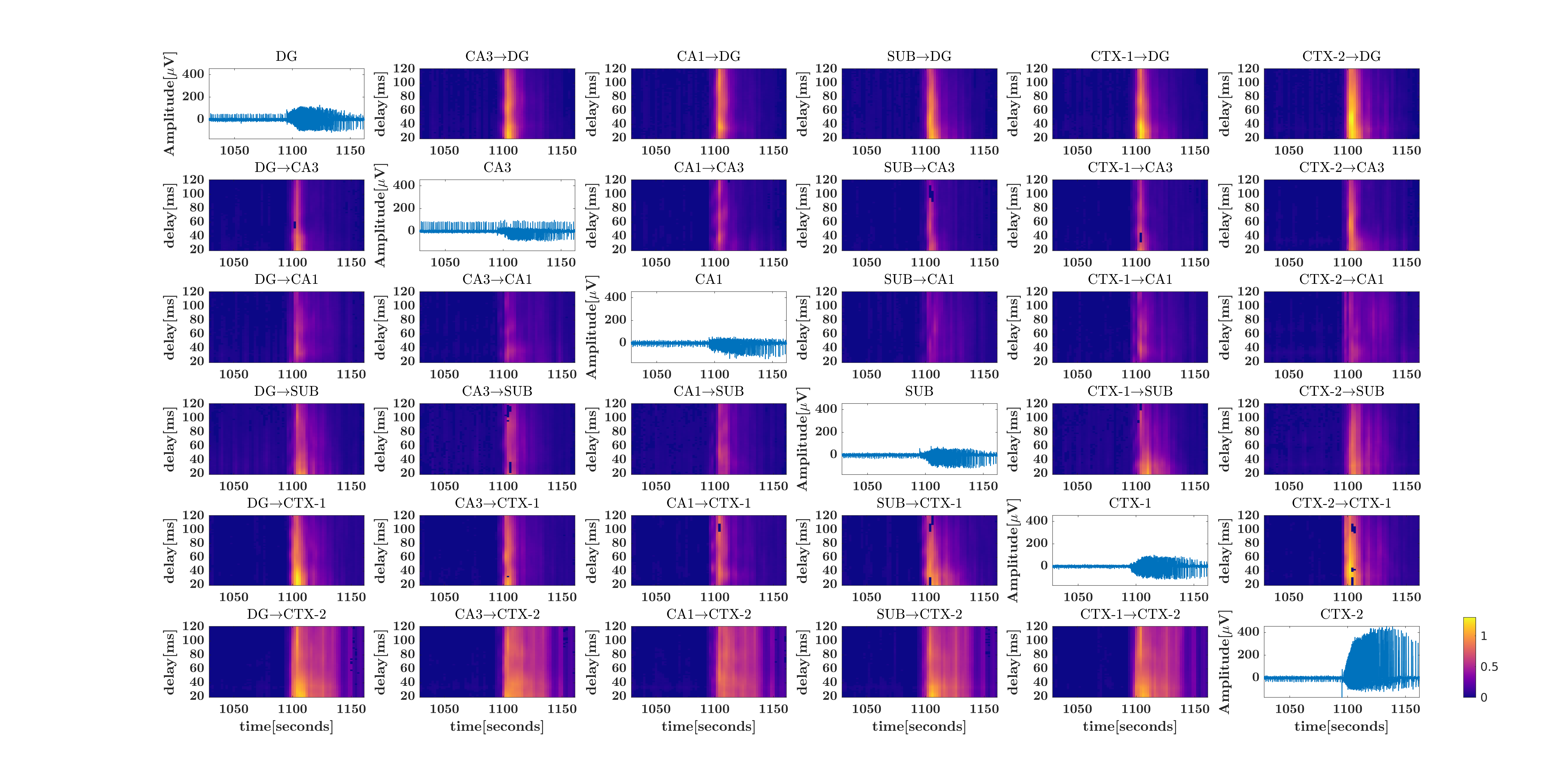}
    \caption{Causality detection based on M-OPTNs for MEA data from slice no. 3. Each off-diagonal subplot represents the time-varying interaction as given by CE based on M-OPTNs between two ROIs for varying time delays. The subplots on the diagonal show the MEA signal (including the ictal activity preceded and followed by inter-ictal activity) in the electrodes placed in each of the six ROIs. The color bar indicates the connection strength.}
    \label{mea_3}
\end{figure}

\begin{figure}
    \centering
    \includegraphics[scale=0.28]{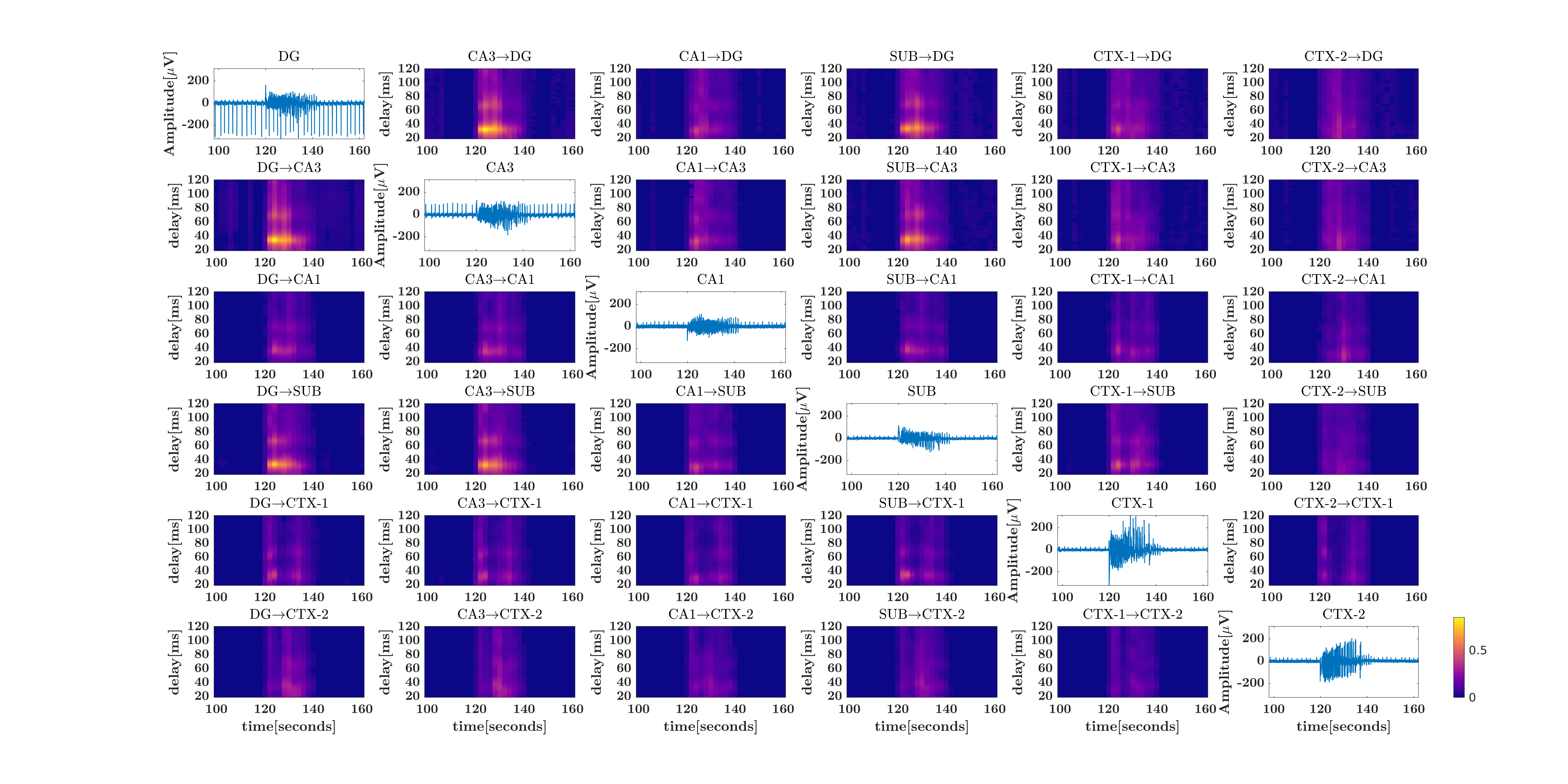}
    \caption{Causality detection based on M-OPTNs for MEA data from slice no. 4. Each off-diagonal subplot represents the time-varying interaction as given by CE based on M-OPTNs between two ROIs for varying time delays. The subplots on the diagonal show the MEA signal (including the ictal activity preceded and followed by inter-ictal activity) in the electrodes placed in each of the six ROIs. The color bar indicates the connection strength.}
    \label{mea_4}
\end{figure}

The results from the entire data sets are qualitatively similar (see Figures~\ref{mea_2}, \ref{mea_3} and \ref{mea_4} in Section \ref{appendix} for the results obtained from the other three brain slices). The propagation from SUB $\rightarrow$ DG was also observed in three out of four slices. The interaction SUB $\rightarrow$ CTX was observed in all slices although the strength of interaction from SUB $\rightarrow$ CTX was found to be generally weaker compared to the interaction CTX $\rightarrow$ SUB. In general, the results show that outward connections from CTX and DG are generally the strongest during the ictal event and that these interactions appear to be strongest at a delay of $\approx 38$ ms.

\section{Discussion}
In this work, we have proposed a new method to detect causality from multivariate observational data by computing information theoretic measures such as CE upon the resulting M-OPTNs. For reliable computation of CE and removal of non-causal neighbors, we have also proposed a pragmatic methodology to define a minimal set for conditioning variables. Our numerical experiments show that our approach can be used to reliably infer both the directionality and the delay of the interactions between the signals even at considerably high $NL$. Causal network inference from real word data of MEA recordings demonstrates that the application of the proposed method can infer network interactions during ictal activity and their associated delays. 

\subsection{Minimal set of neighbors for conditioning}
In order to test if a node (signal) $m$ has a causal influence on node (signal) $n$, the standard and most common approach is to use the Peter and Clarke algorithm \cite{Ref18}, which tests the conditional independence between two nodes given all other variables. Non-existence of a causal relationship between $m$ and $n$ is established once the algorithm finds that $m$ and $n$ are conditionally independent given other variables. When inferring causality on multivariate time series, the conditioning set could have many variables resulting in unreliable estimates of information theoretic measures, particularly when the sample size is small. We alleviate this problem by defining a selection of variables to condition on, based on the common information shared between them. Furthermore, we restrict the number of variables to condition on, $r \le 3$, according to the finite data size ($T \approx 10000$ samples), since conditioning on a higher number of variables resulted in unreliable estimates of CE. In addition to the Peter and Clarke algorithm, there are other iterative, constraint-based approaches that have been proposed for conditioning, including the modified Peter and Clarke algorithm \cite{Ref18} and fast causal inference algorithm \cite{Ref19} as well as score-based algorithms such as greedy equivalence search \cite{Ref20} for defining conditioning sets. However, most of these algorithms suffer from undesirable computational complexity and are not straightforward to implement. A systematic comparison of various conditioning approaches is beyond the scope of this work, where the motivation is to propose and demonstrate M-OPTNs as an extension of OPTNs to reliably infer causality among time series.

\subsection{Effect of various parameters}
The numerical results obtained from simulations of coupled stochastic processes and interacting Lorenz systems, have shown that the proposed approach can successfully capture coupling directions and the associated delays. When applied to more realistic simulations using a network of neural mass models, our approach could reliably recover the underlying causal coupling structure. However, as discussed in detail in the following sections, the performance of the proposed method depends on the choice of several parameters, which can overall be categorized into 
\begin{inparaenum}[(i)]
\item number of time samples and
\item threshold for identifying significant connections.
\end{inparaenum}

\subsubsection{Effect of varying the number of time samples}
It is well known that the amount of data required for reliable reconstruction of the attractor depends on the embedding dimension $M$. Since the probability distributions required for the computation of CE are estimated from the ordinal patterns obtained after embedding, inadequate data length might result in unreliable estimates of CE. In our simulations we have set $M=3$ and found that the results are reliable if $T\gtrsim 10^3$. For a given $M$, there are $(M!)^2$ possible pairs of ordinal patterns for which we have to estimate the co-occurrence frequencies. Having $T \leq 10^M$ samples, results in many spurious interactions being classified as causal links, which is reflected as an increase in the $FPR$ and consequently low $F_1$-score as shown in our simulations. For the stochastic model system, interacting Lorenz systems and the network of neural mass models, we used $M=3$. Embedding in a higher dimension, for example $M=5$ would require $T>100000$ samples for reliable embedding and computation of entropy values. The typical sampling frequency of real-world data such as MEA recordings is of the order of $5000$--$10000$~Hz, and to estimate dynamic changes in a causal effect network based on $100000$ samples would mean using a window size of $10$ to $20$ seconds, which may be far too long compared to dynamical changes that occur in neural networks. Also, the use of $100000$ or more samples, increases the computation time drastically. Thus, for electrophysiological recordings from neural data, a window size of $2$ to $4$ seconds seems more realistic, which amounts to having $10000$  to $20000$ samples per window, depending on the sampling frequency. This in turn means that $M$ should not be greater than $3$ or at most $4$. In contrast, we observed that varying the embedding delay $d$ did not affect the results qualitatively (not shown here) and we used $d=100$ for all our simulations and experimental data.

\subsubsection{Effect of varying threshold parameters to define significant connections}
The parameter $\lambda$ determines the connections (direct and indirect) to be classified as significant. Lower values of $\lambda$ prune away most of the connections, whereas higher values retain most of the connections. Based on our simulations of network of neural mass models, we observed that when the network connectivity is less than $25\%$ and only a moderate amount of observation noise ($NL=0.50$) is present, the choice of $\lambda \geq 0.99$ seems to be the optimal setting that results in high $TPR$ and low $FPR$ and, consequently, high $F_1$-score , provided that $\delta$ is chosen in the range $0.08 \leq \delta \leq 0.12$. However when the $NL$ increases ($NL>0.5$), a setting of $\lambda>0.995$ and $0.08 < \delta < 0.1$ leads to an optimal performance of our algorithm. 

The aforesaid implies that the choice of $\delta$, which determines the threshold to distinguish a causal neighbor from a non-causal neighbor, depends on the amount of noise present in the data. In the presence of low or moderate observational noise, a truly causal neighbor would result in a high $\delta$ as conditioning on this neighbor should reduce the entropy significantly. In contrast, a non-causal neighbor would result in a very small $\delta$. Our results show that the setting $\delta \leq 0.1$ in such a scenario is a reasonable choice along with $\lambda \approx 0.995$. If the data is very noisy ($NL\geq 1$), then the necessary $\delta$ for identifying a truly causal neighbor would be very small, thereby making it hard to distinguish from a non-causal neighbour, for which $\delta$ should also be small. Thus setting $\delta$ too high will prune away all the true connections along with the spurious ones, while setting $\delta$ too low might retain some spurious connections. 

Another factor that impacts the choice of $\lambda$ and $\delta$ in addition to $NL$ is the number of connections in the network. When the network is densely (in case of our simulations, more than $25\%$) connected, finding an optimal $\delta$ and $\lambda$ that gives high $TPR$ and low $FPR$, and consequently a high $F_1$-score is more challenging as the estimated network has many spurious connections at multiple delays in addition to the interactions at the correct delays. Any choice of $\delta$ to prune away these spurious connections will also yield the removal of true connections as the $NL$ increases.

In summary, our results indicate that setting $\lambda \approx 0.995$ and varying $\delta$ between $0.05$ and $0.1$ should result in reliable network inference, assuming the underlying networks are sparse (see Figures~\ref{tpr_NMM} --~\ref{f1_NMM}). 

\subsection{Causal network inference from MEA data}
As a real-world example, we have applied the proposed method to MEA electrophysiology recordings obtained from an in vitro model of acute limbic seizures. Specifically, we have used 4AP-treated rodent hippocampus-CTX slices as a simplified model of the primary neural circuits involved in temporal lobe epilepsy. This model has been extensively characterized \cite{Ref21} and provides a solid ground-truth to validate our approach.

In a previous study \cite{Ref21}, the reported mean time delay for the ictal discharge propagation in the directions CTX $\rightarrow$ DG, CTX $\rightarrow$ CA3 and CTX $\rightarrow$ CA1 was $37.5 \pm 9.6$ ms, $71.7 \pm 27.5$ and $31 \pm 6.3$ ms. In keeping with this, our method has detected connectivity in the direction of CTX $\rightarrow$ DG, CTX $\rightarrow$ CA3 and CTX $\rightarrow$ CA1 at delays of $30-38$ ms during the ictal discharge. Note that the short delay in the CTX $\rightarrow$ CA1 direction is due to the signal propagation along the direct temporoammonic pathway \cite{Ref21}, which is known to short-circuit the hippocampal loop. Moreover, the connection between SUB and CTX is consistent with the previously reported role of SUB-CTX interactions through the temporoammonic pathway in reinforcing ictal synchronization in animal models of temporal lobe epilepsy \cite{Panuccio2010}.

We also found connections in the direction DG $\rightarrow$ CTX, CA3 $\rightarrow$ CTX, CA3 $\rightarrow$ SUB and DG $\rightarrow$ SUB. However, as these connections are disrupted by the Schaffer Collaterals cut, they represent false positives due to far-field contamination \cite{Inaba2006} of the signals recorded from CA1, SUB and CTX, wherein far fields originating in CA3 can be seen in Figure~\ref{ex_mea}. In keeping with this, we found that the strength of the false positive connection is generally lower than the expected true connections. The observation of such false positive interactions could also stem from the common driver issue (see Figure~\ref{example_networks} (B)) wherein one of the CTX ROI is driving both DG and CTX (in the other ROI) causing a spurious DG $\rightarrow$ CTX connection.

Overall, these results support the reliability and usefulness of our approach for analyzing interactions and their delays in real-world observations, such as electrophysiological time series.

\subsection{Future work and perspectives}
Although the proposed algorithm can reliably perform causal inference, there are certain issues that warrant our attention. First, we have not compared our method to other existing techniques that are commonly used to infer causality from electrophysiological recordings. In fact, the main motivation behind this work was to introduce and provide a proof-of-concept that complex network based time series analysis methods such as OPTNs can be generalized and improved to detect causality from multivariate observations. A systematic comparison of our approach with other commonly used techniques to infer causality is beyond the scope of this paper and will remain a subject of future studies. 

Second, we applied our method to \textit{in vitro} MEA electrophysiology data, which is not as widely used to map connections among brain regions  as $\textit{in vivo}$ recordings. In the case of EEG data we typically do not have the ground-truth to validate our method against. In the case of the MEA data used in this study, previous studies have described the anatomical and functional circuits associated within this brain slice preparation, which served as the reference for our results. Furthermore, estimating causality directly from EEG recordings is not trivial due to the issue of volume conduction. In the case of EEG recordings, to mitigate the volume conduction effects, connectivity is estimated from source time-series obtained after solving the EEG inverse problem \cite{Ref44}. Thus, the proposed causal inference method has to be applied on the inverse solution, rather than directly to the EEG data, for reliable causal inference. It is not yet clear how such a transformation would alter the structural properties of the time series, and which impact it could have on the estimation of the ordinal patterns remains as a subject of future studies.  

Third, we did not perform any surrogate data testing but rather relied on the theoretical maximum $H_{max} = \log_2 M!$ and used $\lambda H_{max}$ as threshold for identifying significant bivariate connections, where $0.99 \leq \lambda <1$, as due to the finite sample-size, two independent processes will not have a CE value exactly equal to $H_{max}$. We found this approach to be much faster than generating bivariate surrogates that gave essentially similar results (not shown). In STEP VI of the proposed algorithm, we use the threshold $\delta$ to distinguish between causal and non-causal neighbors. This step can be considerably improved by performing significance testing for the conditional independence test as proposed in \cite{Ref1}, which preserves, for example, the association between $X$ and $Y$ in the coupling scheme $X \leftarrow Z \rightarrow Y$, that would otherwise be destroyed in a strictly bivariate permutation scheme.

\section{Conclusions}
In this paper, we have developed a new method based on OPTNs to infer causality from multivariate observational data. The proposed method allows to infer causality at different delays and can be adapted to provide a time-varying measure of causality. We have also proposed an iterative scheme to find a minimal set of neighbors for conditioning to yield a reliable estimation of the co-occurrence entropy as the employed coupling indicator. We have demonstrated the validity of our approach using different types of simulated signals as well as real-world electrophysiological time series. 

In conclusion, the proposed approach provides a complementary tool for detecting causality from multivariate time series data and can be particularly useful in the area of neuroscience, where the estimation of (time-varying) causal networks from electrophysiology recordings has remained a fundamental problem so far.

\label{sec:4}
\section*{Conflict of Interest} The authors declare that they have no conflict of interest.

\section{Acknowledgements}
This project has received funding from the European Union’s Horizon 2020 research and innovation programme FETPROACT-01-2018 (RIA) awarded to the project Hybrid Enhanced Regenerative Medicine Systems (HERMES) under grant agreement No 824164.

\section{Data availability}
The implementation of the causal inference algorithm and the datasets generated and/or analysed during the current study are available at \url{https://github.com/narayanps/causal_inference_with_OPTNs}.

\section{Appendix}
\label{appendix}
In the neural mass model studied in Section~\ref{sec:3.3}, the neuronal activity in one region is represented by the following set of ordinary differential equations:
\subsubsection*{Pyramidal neurons}
\begin{align}
    \frac{dy_p(t)}{dt} &= x_p(t), \\
    \frac{x_p(t)}{dt} &= G_eh_ez_p(t) - 2h_ex_p(t) - h_e^2y_p(t) \\
    z_p(t) &= \frac{2e_0}{1 + e^{-rv_p}} - e_0 \\
    v_p(t) &= C_{pe}y_e(t) - C_{ps}y_s(t) - C_{pf}y_f(t)
\end{align}
\subsubsection*{Excitatory interneurons}
\begin{align}
    \frac{dy_e(t)}{dt} &= x_e(t), \\
    \frac{x_e(t)}{dt} &= G_eh_e(z_e(t) + \frac{u_p(t)}{C_{pe}}) - 2h_ex_e(t) - h_e^2y_e(t) \\
    z_e(t) &= \frac{2e_0}{1 + e^{-rv_e}} - e_0 \\
    v_e(t) &= C_{ep}y_p(t)
\end{align}
\subsubsection*{Slow inhibitory interneurons}
\begin{align}
    \frac{dy_s(t)}{dt} &= x_s(t), \\
    \frac{x_s(t)}{dt} &= G_sh_sz_s(t) - 2h_sx_s(t) - h_s^2y_s(t) \\
    z_s(t) &= \frac{2e_0}{1 + e^{-rv_s}} - e_0 \\
    v_s(t) &= C_{sp}y_p(t)
\end{align}
\subsubsection*{Fast inhibitory interneurons}
\begin{align}
    \frac{dy_f(t)}{dt} &= x_f(t), \\
    \frac{x_f(t)}{dt} &= G_fh_fz_f(t) - 2h_fx_f(t) - h_f^2y_f(t), \\
    \frac{dy_f(t)}{dt} &= x_l(t), \\
    \frac{x_l(t)}{dt} &= G_eh_eu_f(t) - 2h_ex_l(t) - h_e^2y_l(t), \\
    z_f(t) &= \frac{2e_0}{1 + e^{-rv_f}} - e_0 \\
    v_f(t) &= C_{fp}y_p(t) - C_{fs}y_s(t) - C_{ff}y_l(t)
\end{align}

A network of neural mass models can be constructed by connecting several such regions using a weight matrix $W$ that describes the strength of connections. For example, if $i$ and $j$ represent two regions of neuronal population, then we can define
\begin{equation}
    u_p^i(t) = n_p^i(t) + W_p^{ij}z_p^j(t-d),
\end{equation}
where $u_p(t)$ and $z_p(t)$ correspond to the input and pulse density of the pyramidal neurons, respectively. Analogous definitions apply to the fast inhibitory interneurons. The term $n_p(t)$ represents Gaussian noise with mean $m=0$ and variance $\sigma^2=5$ and $d$ represents the connection delay. For further description on the model, the reader is kindly referred to \cite{Ref22}.


%
%

\bibliographystyle{spphys}       

\begin{thebibliography}{}
%
%
\bibitem{Ref1}
J. Runge, Causal network reconstruction from time series: From theoretical assumptions to practical estimation, Chaos, 28, 075310 (2018)


\bibitem{Ref26}
K. Hlaváčková-Schindler, M. Paluš., M. Vejmelka, and J. Bhattacharya, Causality detection based on information-theoretic approaches in time series analysis. Phys. Rep., 441, 1-46 (2007).

\bibitem{Ref2}
J. Sun and E.M. Bolt, Causation entropy identifies indirect influences, dominance of neighbors and anticipatory couplings. Physica D, 267, 49-57 (2014)

\bibitem{Ref3}
C.W.J. Granger, Investigating causal relations by econometric models and
cross-spectral methods. Econometrica, 37, 424–438 (1969).

\bibitem{Ref4}
 L.A. Baccala and K. Sameshima, Partial directed coherence: A new concept in neural structure determination. Biol. Cybern., 84, 463–474 (2001).
 
 \bibitem{Ref7}
A. Montalto, L. Faes and D. Marinazzo, MuTE: A MATLAB Toolbox to Compare Established and Novel Estimators of the Multivariate Transfer Entropy. PLoS One, 9, e109462 (2014).

\bibitem{Ref5}
T. Schreiber, Measuring information transfer. Phys. Rev. Lett., 85, 461–464 (2000).

\bibitem{Ref6}
J. A. Vastano and H. L. Swinney, Information transport in spatiotemporal systems. Phys. Rev. Lett., 60, 1773 (1988).


\bibitem{Ref8}
S. Li, Y. Xiao, D. Zhou, and D. Cai , Causal inference in nonlinear systems: Granger causality versus time-delayed mutual information. Phys. Rev. E, 97, 052216 (2018).

\bibitem{Ref9}
L. Barnett,  A.B. Barrett, and A.K. Seth, Granger causality and transfer entropy are equivalent for Gaussian variables. Phys. Rev. Lett., 103, 238701 (2009).

\bibitem{Ref32}
K.J. Friston, Functional and effective connectivity in neuroimaging: a synthesis. Human Brain Mapping, 2, 56-78, (1994)

\bibitem{Ref35}
J.P. Lachaux, E.Rodriguez, J. Martinerie and F.J.Francisco. Measuring phase synchrony in brain signals. Human Brain Mapping. 8, 194-208 (1999).

\bibitem{Ref36}
F. Mormann, K. Lehnertz, P. David and C.E. Elger. Mean phase coherence as a measure for phase synchronization and its application to the EEG of epilepsy patients. Physica D, 144, 358-359 (2000).

\bibitem{Ref37}
R. Srinivasan, P.L.Nunez and R.B. Silberstein. Spatial filtering and neocortical dynamics: estimates of EEG coherence. IEEE Trans. Biomed. Engin., 45, 814-826 (1998) 

\bibitem{Ref38}
J. Jeong, J.C.Gore and B.S. Peterson. Mutual information analysis of the EEG in patients with Alzheimer's disease. Clin. Neurophysiol., 112, 827-835 (2001).

\bibitem{Feldt2011}
S. Feldt, P. Bonifazi, and R. Cossart,  Dissecting functional connectivity of neuronal microcircuits: experimental and theoretical insights, 34(5), 225-236 (2011).

\bibitem{Ref33}
I. M. de Abril, J. Yoshimoto and K. Doya. Connectivity inference from neural recording data: Challenges, mathematical bases and research directions. Neural Networks, 102, 120-137 (2018).

\bibitem{Ref39}
R. Vicente, M. Wibral, M. Linder and G. Pipa. Transfer entropy—a model-free measure of effective connectivity for the neurosciences. J. Comput. Neurosci., 30, 45-67 (2011).

\bibitem{Ref45}
N. P. Subramaniyam, F. Tronarp, S. Särkkä and L. Parkkonen. Expectation–maximization algorithm with a nonlinear Kalman smoother for MEG/EEG connectivity estimation. EMBEC and NBC, IFMBE Proceedings, 65 (2017).

\bibitem{Ref46}
F. Tronarp, N.P. Subramaniyam, S. Särkkä and L. Parkkonen. N.P. Subramaniyam. Tracking of dynamic functional connectivity from MEG data with Kalman filtering. 40th Annual International Conference of the IEEE Engineering in Medicine and Biology Society (EMBC) (2018).

\bibitem{Ref47}
N. P. Subramaniyam, F. Tronarp, S. Särkkä and L. Parkkonen. Joint estimation of neural sources and their functional connections from MEG data. bioRxiv preprint ; \url{https://www.biorxiv.org/content/early/2020/10/05/2020.10.04.325563} (2020)

\bibitem{Ref40}
K.J. Friston, L. Harrison and W. Penny. Dynamic causal modelling. Neuroimage. 19, 1273-1302 (2003).

\bibitem{Ref41}
M. Ding, Y. Chen and S.L. Bressler. Granger causality: basic theory and application to neuroscience. Handbook of Time Series Analysis: Recent Theoretical Developments and Applications, 437-460 (2006).

\bibitem{Ref42}
A.M. Bastod and J.M. Schoffelen. A tutorial review of functional connectivity analysis methods and their interpretational pitfalls. Frontiers in systems neuroscience, 9,175 (2016).

\bibitem{Ref10}
Y. Zou, R.V. Donner, N. Marwan, J.F. Donges and J. Kurths, Complex network approaches to nonlinear time series analysis. Phys. Rep., 787, 1-97 (2019).

\bibitem{Ref11}
R.V. Donner, Y. Zou, J.F. Donges, N. Marwan and J. Kurths,  Recurrence networks—a novel paradigm for nonlinear time series analysis. New J. Phys., 12, 033025 (2010).

\bibitem{Ref27}
N.P. Subramaniyam and J. Hyttinen Characterization of dynamical systems under noise using recurrence networks: Application to simulated and EEG data. Phys. Lett. A, 378, 3464-3474 (2014).

\bibitem{Ref28}
N.P. Subramaniyam, J.F. Donges, and J. Hyttinen, Signatures of chaotic and stochastic dynamics uncovered with $\epsilon$-recurrence networks. Proc. R. Soc. A, 471(2183), 20150349 (2015).


\bibitem{Ref12}
L. Lacasa, B. Luque, F. Ballesteros, J. Luque, and J. C. Nuno, From time series to complex networks: The visibility graph. Proc. Natl. Acad. Sci. USA 105(13), 4972 (2008).

\bibitem{Ref13}
M. McCullough, M. Small, T. Stemler and  H.H.C. Iu, Time lagged ordinal partition networks for capturing dynamics of continuous dynamical systems. Chaos, 25, 053101 (2015).

\bibitem{Ref14}
Y. Ruan,  R.V. Donner, S. Guan and Y. Zou, Ordinal partition transition network based complexity measures for inferring coupling direction and delay from time series. Chaos, 29, 043111 (2019).

\bibitem{Ref30}
M. McCullough M, K. Sakellariou , T. Stemler , M. Small, Counting forbidden patterns in irregularly sampled time series. I. The effects of under-sampling, random depletion, and timing jitter. Chaos, 26, 123103 (2016).

\bibitem{Ref29}
C. Bandt and B. Pompe, Permutation entropy: a natural complexity measure for time series. Phys. Rev. Lett., 88, 174102 (2002).

\bibitem{Ref17}
J. Y. Zhang, J. Zhou, M. Tang, H. Guo, M. Small, and Y. Zou, Constructing ordinal partition transition networks from multivariate time series. Sci. Rep. 7, 7795 (2017).

\bibitem{Amigo2008}
J. M. Amig\'o, S. Zambrano, and M. A. F. Sanju\'an, Combinatorial detection of determinism in noisy time series. Europhys. Lett., 83, 60005 (2008).

\bibitem{Amigo2010}
J. M. Amig\'o, S. Zambrano, and M. A. F. Sanju\'an, Detecting determinism in time series with ordinal patterns: a comparative study. Int. J. Bifurcation Chaos, 20, 2915 (2010).

\bibitem{Kulp2016a}
C. W. Kulp, J. M. Chobot, B. J. Niskala, and C. J. Needhammer, Using forbidden ordinal patterns to detect determinism in irregular sampled time series. Chaos, 26, 023107 (2016).

\bibitem{Ref43}
J.M. Amigo, K.Keller and V.A. Unakafova. Ordinal symbolic analysis and its application to biomedical recordings. Phil. Trans. R. Soc. A, 373, 20140091 (2015).

\bibitem{Ref15}
C. W. Kulp, J. M. Chobot, H. R. Freitas, and G. D. Sprechini, Using ordinal partition transition networks to analyze ECG data. Chaos, 26, 073114 (2016).

\bibitem{Ref16}
K. Keller, A.M. Unakafov, and V.A. Unakafova, Ordinal patterns, entropy, and EEG. Entropy, 16, 6212-6239 (2014).

\bibitem{Ref22}
M. Zavaglia, F. Cona and M. Ursino, A Neural Mass Model to Simulate Different Rhythms in a Cortical Region. Comput. Intell. Neurosci., 2010, 456140 (2010).

\bibitem{Ref25}
G. Panuccio, I. Colombi, and M. Chiappalone, Recording and Modulation of Epileptiform Activity in Rodent Brain Slices Coupled to Micro Electrode Arrays. J. Vis. Exp., 135, e57548 (2018).

\bibitem{Ref31}
P. A. Rutecki, F. J. Lebeda, and D. Johnston, 4-Aminopyridine produces epileptiform activity in hippocampus and enhances synaptic excitation and inhibition. J. Neurophysiol., 57, 1911-1924 (1987).

\bibitem{Ref23}
D. G. Amaral and M. P. Witter, The three-dimensional organization of the hippocampal formation: a review of anatomical data. Neuroscience, 31, 571-591 (1989).

\bibitem{Ref24}
M. Avoli, M. D’Antuono, J. Louvel, R. K\"ohling, G. Biagini, R. Pumain, G. D’Arcangelo, and V. Tancredi, Network and pharmacological mechanisms leading to epileptiform synchronization in the limbic system in vitro. Prog. Neurobiol., 68, 167-207 (2002).

\bibitem{Ref21}
M. Barbarosie, J.Louvel, I.Kurcewicz and M. Avoli, CA3-released entorhinal seizures disclose dentate gyrus epileptogenicity and unmask a temporoammonic pathway. J. Neurophysiol., 83, 1115-1124 (2000).


\bibitem{Ref18}
P. Spirtes, C.N. Glymour, R. Scheines, and D. Heckerman, Causation, prediction, and search. MIT Press, Boston (2000).

\bibitem{Ref19}
J. Ramsey, J. Zhang, and P. L. Sprites, Adjacency-faithfulness and conservative causal inference. arXiv preprint arXiv:1206.6843 (2012).

\bibitem{Ref20}
D. M. Chickering, Learning equivalence classes of Bayesian-network structures. J. Machine Learn. Res., 2, 445-498 (2002).

\bibitem{Panuccio2010}
G. Panuccio, M. D'Antuono, P. De Guzman, L. De Lannoy, G. Biagini and M. Avoli. In vitro ictogenesis and parahippocampal networks in a rodent model of temporal lobe epilepsy. Neurobiol. Disease, 39, 372-380 (2010).

\bibitem{Inaba2006}
Y.Inaba and M.Avoli. Volume-conducted epileptiform events between adjacent necortical slices in an interface tissue chamber. Journal of Neuroscience Methods, 151, 287-290 (2006). 
\bibitem{Ref44}
C. Brunner, M. Billinger, M. Seeber, T.R.Mullen and S.Makeig. Volume conduction influences scalp-based connectivity estimates. Front. Comput. Neurosci., 10, 121 (2016).

\end{thebibliography}


\end{document}